\documentclass[preprint,12pt]{elsarticle}

\usepackage{amssymb}
\usepackage{amsmath}
\usepackage{lineno}

\biboptions{compress}

\usepackage[table]{xcolor}
\usepackage{multirow}
\usepackage[scriptsize,tight,figbotcap]{subfigure}
\usepackage[cm]{fullpage}

\usepackage{bm}   % bold mathematics fonts, including greek and other symbols

\usepackage{doi}
\usepackage{hyperref}
\hypersetup{
    colorlinks,
    linkcolor={red!50!black},
    citecolor={blue!50!black},
    urlcolor={blue!80!black}
}

\journal{Journal of Non-Newtonian Fluid Mechanics}

\begin{document}
\newcommand{\vf}[1]{\bm{\mathrm{#1}}}
\newcommand{\pd}[2]{\frac{\partial #1}{\partial #2}}

\begin{frontmatter}



\title{Solution of the square lid-driven cavity flow of a Bingham plastic using the finite volume method}


\author[oc]{Alexandros Syrakos}
\ead{syrakos.alexandros@ucy.ac.cy}

\author[dms,oc]{Georgios C. Georgiou\corref{cor1}}
\ead{georgios@ucy.ac.cy}

\author[dmme]{Andreas N. Alexandrou}
\ead{andalexa@ucy.ac.cy}

\cortext[cor1]{Corresponding author}

\address[oc]{Oceanography Centre, University of Cyprus, PO Box 20537, 1678 Nicosia, Cyprus}
\address[dms]{Department of Mathematics and Statistics, University of Cyprus, PO Box 20537, 1678 Nicosia, Cyprus}
\address[dmme]{Department of Mechanical and Manufacturing Engineering, University of Cyprus, PO Box 20537, 1678 Nicosia,
Cyprus}

\begin{abstract}
We investigate the performance of the finite volume method in solving viscoplastic flows. The creeping square lid-driven
cavity flow of a Bingham plastic is chosen as the test case and the constitutive equation is regularised as proposed by
Papanastasiou [J. Rheology 31 (1987) 385-404]. It is shown that the convergence rate of the standard SIMPLE
pressure-correction algorithm, which is used to solve the algebraic equation system that is produced by the finite
volume discretisation, severely deteriorates as the Bingham number increases, with a corresponding increase in the
non-linearity of the equations. It is shown that using the SIMPLE algorithm in a multigrid context dramatically improves
convergence, although the multigrid convergence rates are much worse than for Newtonian flows. The numerical results
obtained for Bingham numbers as high as 1000 compare favorably with reported results of other methods.
\end{abstract}

\begin{keyword}
Bingham plastic \sep Papanastasiou regularisation \sep lid-driven cavity \sep finite volume method \sep SIMPLE \sep
multigrid

\end{keyword}

\end{frontmatter}

This is the accepted version of the article published in: Journal of Non-Newtonian Fluid Mechanics 195 (2013) 19--31, 
\doi{10.1016/j.jnnfm.2012.12.008}

\textcopyright 2016. This manuscript version is made available under the CC-BY-NC-ND 4.0 license 
\url{http://creativecommons.org/licenses/by-nc-nd/4.0/}

% \begin{linenumbers}

\section{Introduction}
\label{sec: introduction}

Viscoplastic flows constitute an important branch of non-Newtonian fluid mechanics, as many materials of industrial, geophysical, 
and biological importance are known to exhibit yield stress. In general, yield-stress fluids are suspensions of particles or 
macromolecules, such as pastes, gels, foams, drilling fluids, food products, and nanocomposites. A comprehensive review of 
viscoplasticity has been carried out by Barnes \cite{Barnes_99}. Such materials behave as (elastic or inelastic) solids, below a 
certain critical shear stress level, i.e. the yield stress, and as liquids otherwise. The flow field is thus divided into 
unyielded (rigid) and yielded (fluid) regions. 

The simplest constitutive equation describing viscoplasticity is that proposed by Bingham \cite{Bingham_22}:

\begin{equation} \label{eq: Bingham_constitutive}
 \left\{ \begin{array}{ll} 
   \vf{\dot{\gamma}} \;=\; \vf{0} \;, \qquad  &  \tau \leq \tau_y
\\
   \vf{\tau} \;=\; \left( \dfrac{\tau_y}{\dot{\gamma}} + \mu \right)\vf{\dot{\gamma}} \;, & \tau > \tau_y
 \end{array} \right.
\end{equation}
where $\tau_y$ is the yield stress, $\mu$ is the plastic viscosity, $\vf{\tau}$ is the stress tensor, $\vf{\dot{\gamma}}$ is the 
rate of strain tensor,

\begin{equation} \label{eq: gamma_tensor}
 \vf{\dot{\gamma}} \;\equiv\; \nabla\vf{u} + \left(\nabla\vf{u}\right)^{\mathrm{T}}
\end{equation}
$\vf{u}$ is the velocity vector, and the superscript T denotes the transpose of the velocity-gradient tensor $\nabla\vf{u}$. The 
symbols $\tau$ and $\dot{\gamma}$ denote the magnitudes of the stress and rate-of-strain tensors, respectively:

\begin{equation} \label{eq: tau,gamma invariants}
 \tau         \;\equiv\; \left[ \tfrac{1}{2} \vf{\tau} : \vf{\tau} \right]^{\frac{1}{2}} \quad , \quad
 \dot{\gamma} \;\equiv\; \left[ \tfrac{1}{2} \vf{\dot{\gamma}} : \vf{\dot{\gamma}} \right]^{\frac{1}{2}}
\end{equation}
Another two-parameter viscoplastic equation is the Casson model, which is mostly used in hemodynamics. The Herschel--Bulkley 
model is the generalisation of the Bingham-plastic equation, which involves a power-law exponent allowing shear-thinning or shear 
thickening.

Simulating viscoplastic flows poses extra-ordinary difficulties due to the discontinuity of the constitutive equations. In most 
cases, it is necessary to determine the location and the shape of the yield surface at which the flow switches from one branch of 
the constitutive equation to the other, e.g. from solid to liquid behavior. Mathematically, the yield surface is the locus of 
points where $\tau=\tau_y$. A common approach to overcoming this burden is to regularise the constitutive equation, i.e. to 
describe the two branches of (\ref{eq: Bingham_constitutive}) by one smooth equation using a stress-growth parameter. The most 
popular regularization is that proposed by Papanastasiou \cite{papanastasiou_87}:

\begin{equation} \label{eq: papanastasiou_tau}
 \vf{\tau} \;=\; \left[ \frac{\tau_y}{\dot{\gamma}} \{ 1-\exp(-m\dot{\gamma})\} \,+\, \mu \right] \vf{\dot{\gamma}}
\end{equation}
where $m$ denotes the stress growth parameter, which needs to be ``sufficiently'' large. Frigaard and Nouar \cite{Frigaard_05} 
reviewed systematically the convergence of the Papanastasiou and other regularized models including the bi-viscosity model 
\cite{Donovan_84}.

Another approach in solving viscoplastic flows is based on the use of variational inequalities, that is rate-of-strain 
minimization or stress maximization, which form the basis of the Augmented Lagrangian Method \cite{Fortin_83, Glowinski_84}. Dean 
et al. \cite{Dean_07} reviewed and compared numerical methods based on the variational inequality approach.

In this paper, we adopt the regularisation approach and investigate the performance of the finite volume/multigrid method in 
solving Bingham plastic flows. The finite volume method (FVM) is a popular method for solving fluid flows, employed by many 
general-purpose CFD solvers. One of the attractive features of the method is that it can be applied to a variety of physical 
problems with relative ease. However, there are a limited number of published results on the use of the finite-volume method to 
solve Bingham flow problems. Neofytou \cite{neofytou_05} used a FVM in conjunction with the SIMPLE algebraic solver (Patankar and 
Spalding \cite{Patankar_72}) to simulate the lid-driven cavity flow of various non-Newtonian fluids, including a 
Papanastasiou-regularised Bingham plastic at quite low Bingham numbers (0.01--1). Turan and co-workers \cite{Turan_10, Turan_12} 
also used a commercial FVM/SIMPLE code employing the bi-viscosity model in order to simulate natural convection of a Bingham 
plastic in a square cavity. The FVM was also used to solve flows of a Casson fluid through a stenosis (Neofytou and Drikakis 
\cite{Neofytou_03}) and through a sudden expansion (Neofytou and Drikakis \cite{Neofytou_03b}), at rather low yield-stress 
values. Also, de Souza Mendes et al. \cite{deSouzaMendes_07} and Naccache and Barbosa \cite{Naccache_07} used the FVM in order to 
simulate viscoplastic flow through an expansion followed by a contraction.

A benchmark problem for testing numerical methods for both Newtonian and non-Newtonian flows is the lid-driven cavity flow 
problem. For laminar Newtonian flow, the reader is referred to the works of Botella and Peyret \cite{Botella_98}, Syrakos and 
Goulas \cite{syrakos_06b}, Bruneau and Saad \cite{Bruneau_06}, and the references therein. The lid-driven cavity flow has also 
been used as a test case for Bingham flows by Sanchez \cite{Sanchez_98}, Mitsoulis and Zisis \cite{mitsoulis_01}, Vola et al. 
\cite{Vola_03}, Elias et al. \cite{Elias_06}, Yu and Wachs \cite{Yu_07}, Olshanskii \cite{Olshanskii_09}, Zhang \cite{Zhang_10}, 
and dos Santos et al. \cite{dosSantos_11}, who used solution methods other than the FVM, mostly the Finite Element method. To the 
authors' knowledge, only Neofytou \cite{neofytou_05} used the FVM in order to solve the driven cavity flow of a Bingham plastic 
or any other viscoplastic fluid, albeit for very small Bingham numbers ($\leq 1$). Although some of the aforementioned published 
works contain results for non-zero Reynolds numbers, in most cases the results concern creeping flows ($Re = 0$).

The objective of the present work is to apply the FVM along with a multigrid method in order to improve the convergence of the 
SIMPLE solver when solving the lid-driven cavity Bingham flow for a broad range of Bingham numbers. Multigrid methods use a 
hierarchy of grids of progressive fineness. By applying the algebraic solver on each of these grids, all wavelengths of the 
algebraic error are reduced with equal efficiency. Multigrid methods were originally proposed by Fedorenko \cite{Fedorenko_62} 
and later developed by Brandt \cite{Brandt_77}. In the context of the FVM, they were first used in the late 80's (Sivaloganathan 
and Shaw \cite{Sivaloganathan_88}; Hortman et al. \cite{Hortmann_90}). Since then, they have been employed in numerous studies 
involving the FVM, many of which used the lid-driven cavity flow of a Newtonian fluid as a test problem; see Syrakos and Goulas 
\cite{syrakos_06b} and references therein. These Newtonian studies have shown that using multigrid algorithms can result in very 
significant performance gains. To our knowledge, the finite volume/multigrid method has not been tested in the case of Bingham 
flow.

The rest of the paper is organized as follows. In Section 2, the integral forms of the governing equations are presented and 
dedimensionalized. In Section 3, the numerical method, that is the discretisation of the governing equations and the solution of 
the resulting algebraic system, is discussed. The numerical results are presented in Section 4. These compare well with the 
results of Mitsoulis and Zisis \cite{mitsoulis_01} and Yu and Wachs \cite{Yu_07}. The convergence of the algebraic solver has 
also been studied and possible ways for its acceleration have been investigated. Finally, Section 5 contains our concluding 
remarks.

\section{Governing equations}
\label{sec: equations}

We consider the steady-state, two-dimensional flow in a square cavity of side $L$, the top boundary (lid) of which moves towards 
the right with a uniform horizontal velocity $U$, while the remaining sides are fixed. We work in Cartesian coordinates 
($x$,$y$), centered at the lower-left corner of the cavity and denote the unit vectors in the $x$ and $y$ directions by $\vf{i}$ 
and $\vf{j}$, respectively. Let also $\rho$ and $\eta=\eta(\dot{\gamma})$ denote the density and the viscosity of any 
generalised-Newtonian fluid. By means of the Gauss theorem, the integral forms of the continuity and the $x$-- and 
$y$--components of the momentum equations over a control volume $P$ are as follows:

\begin{equation} \label{eq: continuity}
 \int_{S_P}\! \rho \vf{u} \cdot d\vf{S} \;=\; 0
\end{equation}

\begin{equation} \label{eq: momentum_x}
 \int_{S_P}\! \rho\, u\, \vf{u} \cdot d\vf{S} \;=\; 
 \int_{S_P}\! \eta \left( \nabla u \,+\, \pd{\vf{u}}{x} \right) \!\cdot d\vf{S} \;-\;
 \int_{S_P} p \vf{i} \cdot d\vf{S}
\end{equation}

\begin{equation} \label{eq: momentum_y}
 \int_{S_P}\! \rho\, v\, \vf{u} \cdot d\vf{S} \;=\; 
 \int_{S_P}\! \eta \left( \nabla v \,+\, \pd{\vf{u}}{y} \right) \!\cdot d\vf{S} \;-\;
 \int_{S_P} p \vf{j} \cdot d\vf{S}
\end{equation}
where $S_P$ is the boundary surface of the control volume, $d\vf{S}$ is an infinitesimal element of this surface oriented so that 
the normal vector points out of $P$, $\vf{u}=u\vf{i}+v\vf{j}$ is the velocity vector, and $p$ is the pressure. In the case of the 
Papanastasiou model (\ref{eq: papanastasiou_tau}), the viscosity is given by 

\begin{equation} \label{eq: papanastasiou_eta}
 \eta \;=\; \frac{\tau_y}{\dot{\gamma}}\left\{ 1-\exp(-m\dot{\gamma}) \right\} \:+\: \mu
\end{equation}

The advantage of the Papanastasiou regularization is that expression (\ref{eq: papanastasiou_eta}) is used over the entire flow 
domain, i.e. over both yielded and unyielded regions. At high strain rates the viscosity tends towards $\mu$ if the growth 
parameter $m$ is large enough. In the unyielded regions the viscosity obtains high values which result in very small values of 
$\dot{\gamma}$ and thus solid body motion is approximated. When $\dot{\gamma}$ tends to zero, then the viscosity (\ref{eq: 
papanastasiou_eta}) tends not to infinity but to the finite value $m\tau_y + \mu$. Some authors \cite{tsamopoulos_96, burgos_99} 
suggested that lower values of $m$ can be used at higher yield stress values and vice versa. 

To dedimensionalise the governing equations (\ref{eq: continuity})--(\ref{eq: momentum_y}), we scale lengths by the cavity side 
$L$, the velocity components by $U$, and the pressure and stress by $\mu U/L$, and use stars to denote the dimensionless 
variables. Taking into account that $\rho$ is constant, the dimensionless equations are as follows:

\begin{equation} \label{eq: continuity nd}
 \int_{S_P}\! \vf{u}^* \cdot d\vf{S}^* \;=\; 0
\end{equation}

\begin{equation} \label{eq: momentum_x nd}
 Re \: \int_{S^*_P}\! u^* \, \vf{u}^* \cdot d\vf{S}^* \;=\; 
 \int_{S^*_P}\! \eta^* \left( \nabla^* u^* \,+\, \pd{\vf{u}^*}{x^*} \right) \!\cdot d\vf{S}^* \;-\;
 \int_{S^*_P} p^* \vf{i} \cdot d\vf{S}^*
\end{equation}

\begin{equation} \label{eq: momentum_y nd}
 Re \: \int_{S^*_P}\! v^* \, \vf{u}^* \cdot d\vf{S}^* \;=\; 
 \int_{S^*_P}\! \eta^* \left( \nabla^* v^* \,+\, \pd{\vf{u}^*}{y^*} \right) \!\cdot d\vf{S}^* \;-\;
 \int_{S^*_P} p^* \vf{j} \cdot d\vf{S}^*
\end{equation}

\begin{equation} \label{eq: papanastasiou_eta nd}
 \eta^* \;=\; \frac{Bn}{\dot{\gamma}^*}\left\{ 1-\exp(-M\dot{\gamma}^*) \right\} \:+\: 1
\end{equation}
where
\begin{equation} \label{eq: Re}
 Re \;\equiv \; \frac{\rho U L}{\mu}
\end{equation}
is the Reynolds number,
\begin{equation} \label{eq: Bn}
 Bn \;\equiv\; \frac{\tau_y L}{\mu U}
\end{equation}
is the Bingham number, and
\begin{equation} \label{eq: M}
 M \;\equiv\; \frac{m U}{L}
\end{equation}
is the stress-growth number.

For the sake of simplicity, the stars denoting the dimensionless variables are dropped hereafter.

\section{Numerical method}
\label{sec: method}

\subsection{Discretisation of the equations}
\label{sec: discretisation}
The domain is split into a number of control volumes (CVs) using a Cartesian grid of equally spaced horizontal and vertical grid 
lines. For each control volume, the continuity and momentum equations are approximated using algebraic expressions involving the 
values of the unknowns $u$, $v$, $p$ at the centre of that CV and at the centres of neighbouring CVs \cite{Ferziger_02}. The 
computer code which was used for the present study has the ability to use curvilinear grids composed of quadrilateral CVs - see 
\cite{syrakos_06a, syrakos_06b} for details - but in the present section only the simpler form that these discretisation schemes 
acquire when the grid is Cartesian will be presented. Although the present work concerns only creeping flow, the treatment of the 
convection terms of the momentum equations is also described for completeness.

\begin{figure}[htb]
 \centering
 \includegraphics[scale=1.00]{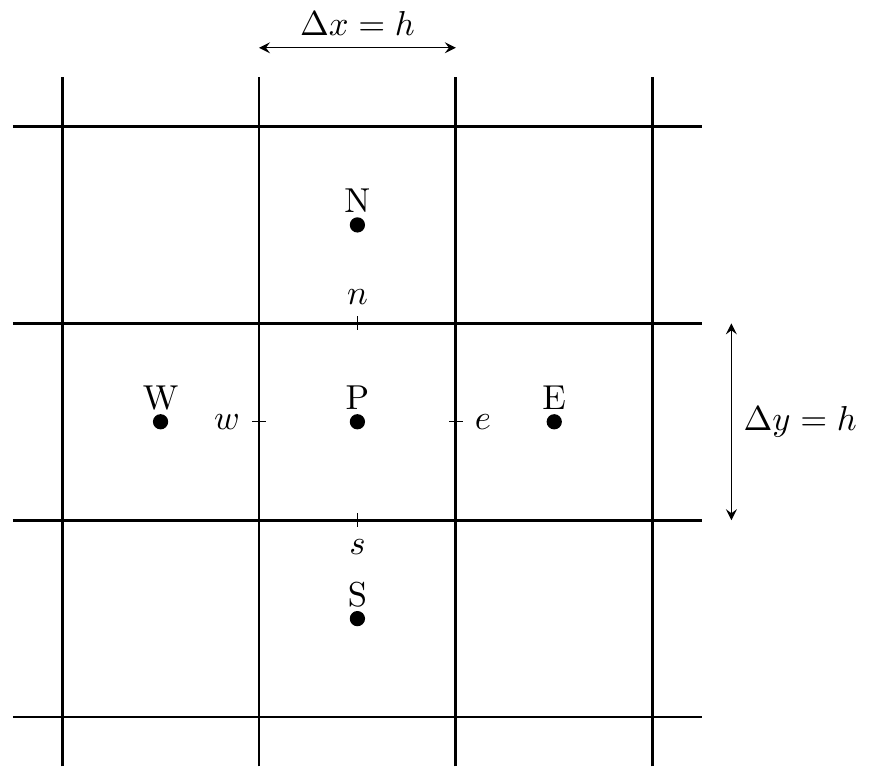}
 \caption{A control volume $P$ and its neighbours.}
 \label{fig: grid}
\end{figure}

For each CV, the surface flux integrals are calculated separately on each face. Figure \ref{fig: grid} shows a control volume $P$ 
and its neighbours, $S$, $E$, $N$ and $W$. The letters $P$, $S$, $E$, $N$ and $W$ will also denote the position vectors of the 
centres of the respective CVs. First, the flow variables and their normal derivatives are calculated at the centre of each face 
using central differences, e.g. for face $e$:

\begin{equation} \label{eq: CDS convection}
 u_e \;=\; \frac{u_E + u_P}{2}
\end{equation}
\begin{equation} \label{eq: CDS diffusion}
 \left. \pd{u}{x} \right|_e \;=\; \frac{u_E - u_P}{\Delta x}
\end{equation}
The flux integrals are then approximated using the midpoint rule, e.g. for the $x$--momentum equation:

\begin{align}
\label{eq: x-mom convection} 
  \int_e\! u\, \vf{u} \cdot d\vf{S} \;&\approx\; F_e u_e
\\%[0.4cm]
\label{eq: x-mom diffusion 1} 
  \int_e\! \eta \nabla u \cdot d\vf{S} \;&\approx\; \eta_e  \left. \pd{u}{x} \right|_e \Delta y
\\%[0.4cm]
\label{eq: x-mom pressure}
  \int_e p\, \vf{i} \cdot d\vf{S} \;&\approx\; p_e \Delta y
\end{align}
Equations (\ref{eq: x-mom convection}) -- (\ref{eq: x-mom diffusion 1}) do not describe the approximation scheme sufficiently, 
and some definitions are still missing: $F_e$ denotes the mass flux through face $e$ and will be defined shortly; the term 
(\ref{eq: x-mom diffusion 1}) is only part of the total viscous flux; and the term $\eta_e$ requires that the viscosity has 
already been calculated somehow at CV centres. To define these terms we first approximate the gradient operator at CV centres 
(equation (\ref{eq: CDS diffusion}) only approximates the normal component at face centres). At the centre of control volume $P$ 
the following approximation is used:

\begin{equation} \label{eq: grad at P}
 \left.\pd{u}{x}\right|_P =\, \frac{u_E - u_W}{2\Delta x} \; , \quad
 \left.\pd{u}{y}\right|_P =\, \frac{u_N - u_S}{2\Delta y}
\end{equation}
If $P$ is a boundary CV, and the domain boundary coincides with face $w$, then the horizontal component of the gradient is 
calculated as follows:

\begin{equation} \label{eq:grad_boundary}
 \left.\pd{u}{x}\right|_P =\, \frac{u_E + u_P - 2u_w}{2\Delta x}
\end{equation}
where $u_w$ is the boundary value of $u$, specified by the Dirichlet boundary condition.

Now, the viscosity is calculated at CV centres from equation (\ref{eq: papanastasiou_eta nd}), where $\dot{\gamma}$ is calculated 
from (\ref{eq: tau,gamma invariants}), using the velocity gradients at CV centres (\ref{eq: grad at P}):

\begin{equation} \label{eq: gamma discretised}
 \dot{\gamma}_P \;=\; \left[ 
   2\left(\left.\pd{u}{x}\right|_P\right)^2  +\,  2\left(\left.\pd{v}{y}\right|_P\right)^2
   +\,  \left( \left.\pd{u}{y}\right|_P + \left.\pd{v}{x}\right|_P \right)^2
 \right]^{\frac{1}{2}}
\end{equation}
Thus $\eta_e$ in (\ref{eq: x-mom diffusion 1}) is calculated from $\eta_P$ and $\eta_E$ using linear interpolation (\ref{eq: CDS 
convection}). Since the viscosity is only needed at face centres, an alternative would be to calculate it there directly. This 
would require the velocity gradients at the face centres, the normal component of which is readily available from (\ref{eq: CDS 
diffusion}). But the tangential component is not, and would require interpolation of velocity components at face vertices. So, 
this alternative would not be less complex than the approach presently adopted.

For the same reason, the remaining part of the viscous fluxes is also calculated by interpolating the velocity gradients at face 
centres from the CV centres. For example, for faces $e$ and $n$ we have, respectively:

\begin{equation}
  \label{eq: x-mom diffusion 2}
  \int_e\! \eta \pd{\vf{u}}{x} \cdot d\vf{S} \;\approx\; \eta_e \left(\pd{u}{x}\right)_e \Delta y
 \; , \quad
  \int_n\! \eta \pd{\vf{u}}{x} \cdot d\vf{S} \;\approx\; \eta_n \left(\pd{v}{x}\right)_n \Delta x
\end{equation}
Here the derivatives $(\partial u/\partial x)_e$ and $(\partial v/\partial n)_e$ are not the same as in Eq. (\ref{eq: CDS 
diffusion}), but are interpolated from the derivatives (\ref{eq: grad at P}) at face centres according to (\ref{eq: CDS 
convection}). This approach is convenient in the case of curvilinear grids, which is why it is adopted by the code we used for 
the present study.

This leaves only the mass fluxes to be defined. These are discretised using the central difference scheme (\ref{eq: CDS 
convection}), but with the addition of an artificial pressure term whose role is to not allow the appearance of spurious pressure 
oscillations in the discrete solution. The mass flux through face $e$, in non-dimensional form, is discretised as:

\begin{equation} \label{eq: F_e}
 F_e \;=\; \int_e \vf{u} \cdot d\vf{S} \;\approx\;
           u_e \Delta y \;+\;
           \frac{(\Delta y)^2}{a_e} \left[ (p_P - p_E) \;-\; \frac{1}{2} \left( \left.\pd{p}{x}\right|_P \!+
           \left.\pd{p}{x}\right|_E \right) (x_P-x_E) \right]
\end{equation}
where:
\begin{equation} \label{eq: a_e}
 a_e \;=\; Re \left( |u_e| \Delta y + |v_e| \Delta x \right) \;+\; 
           2\eta_e \left( \frac{\Delta y}{\Delta x} + \frac{\Delta x}{\Delta y} \right)
\end{equation}
Without this treatment, artificial pressure oscillations do appear when both velocity and pressure are stored at the CV centres, 
as in the present scheme. This technique is known as \textit{momentum interpolation} and was first suggested by Rhie and Chow 
\cite{Rhie_Chow}. The above variant, proposed in \cite{syrakos_06a}, has the advantage that it is decoupled from the SIMPLE 
solution algorithm. The resulting equations simplify further in case of a uniform grid, $\Delta x = \Delta y = h$, but for 
completeness the more general forms are included here. The artificial pressure term is very small, of order $O(h^5)$, and has a 
very small impact on the accuracy of the discretisation of the mass flux.

The test cases examined in this work involve only no-slip solid wall boundaries, with Dirichlet boundary conditions. The 
discretisation there is as follows: The pressure is linearly extrapolated from the interior. The product $\nabla u \cdot d\vf{S}$ 
of the main viscous terms (\ref{eq: x-mom diffusion 1}) is calculated as a one-sided difference, i.e. if $e$ is a boundary face 
then $\nabla u \cdot d\vf{S}$ is approximated as $\Delta y \!\cdot\! (u(e) - u_P)/(\Delta x / 2)$ where $u(e)$ is the exact value 
of $u$ at the centre of face $e$, as defined by the Dirichlet boundary condition. Finally, in the secondary viscous terms 
(\ref{eq: x-mom diffusion 2}) both the derivative and the viscosity at the boundary face centre are taken as equal to their 
values at the adjacent CV centre.

By substituting all terms in the continuity and momentum equations of each CV by their discrete counterparts, a non-linear 
algebraic system is obtained, with four equations ($x$,$y$--momentum, continuity and constitutive equation) and four unknowns 
($u$,$v$,$p$,$\eta$) per CV. The overall discretisation scheme is of second order accuracy, meaning that the discretisation error 
should decrease as $O(h^2)$. Solution of this algebraic system gives the values of the unknowns, up to the discretisation error. 
The solver used to solve this system is described next.

\subsection{Solution of the algebraic system}
\label{sec: solution}

The popular SIMPLE algorithm \cite{Patankar_72} was chosen as the non-linear solver. It is a widely used algorithm and so it will 
not be fully described here -- the interested reader is referred to \cite{Patankar_80} or \cite{Ferziger_02}. In brief, SIMPLE is 
an iterative algorithm which constructs and solves a number of linear systems within each iteration. These linear systems come 
from breaking up and linearising the set of discretised equations: The systems of equations for the momentum components are 
converted into linear systems for the corresponding velocity components by evaluating some of the terms (including the viscosity 
(\ref{eq: papanastasiou_eta nd})) using the velocity and pressure values obtained from the previous SIMPLE iteration; and the 
continuity equations system is used to construct an approximate ``pressure correction'' linear system, which attempts to improve 
the current pressure estimate so as to force the velocity field to be more continuity-conservant. A SIMPLE iteration (outer 
iteration) consists of the successive solution of the linear systems of the velocity components and of pressure correction. Only 
a few iterations (inner iterations) of a linear solver are applied to each linear system, since the matrices of coefficients will 
change in the next outer SIMPLE iteration. In this work we used GMRES as the linear solver for the velocity systems, and 
conjugate gradients (CG) for the pressure correction system, preconditioned by incomplete factorisations with zero fill-in -- see
\cite{Saad} for detailed descriptions. To achieve convergence, underrelaxation factors $a_u, a_p < 1$ are applied to the velocity 
and pressure correction systems respectively \cite{Ferziger_02}. SIMPLE iterations are repeated until the residuals of all the 
original non-linear algebraic equations drop below a selected threshold.

The only modification needed so that SIMPLE can be used for Bingham or other generalised-Newtonian flows is that, at the start of 
every SIMPLE iteration, the viscosity must be updated at each CV according to (\ref{eq: papanastasiou_eta nd}), using the current 
estimate of the velocity field to calculate $\dot{\gamma}$.

The SIMPLE algorithm converges rather slowly, so it was decided to implement it in a multigrid context to accelerate its 
convergence. Many algebraic solvers are able to quickly reduce the short wavelengths of the error but are quite slow in reducing 
the longer wavelengths. SIMPLE is such a solver \cite{Shaw_88}, not only because the linear solvers used for the inner iterations 
may themselves have this property, but also because of local assumptions made in the linearisation of the non-linear terms and in 
the construction of the pressure correction equation, which relate a CV to its direct neighbours. In such a case, after a few 
iterations have been performed and the short wavelengths of the error have been reduced, it is beneficial to move the solution 
procedure to a coarser grid, where the direct neighbours of a CV are farther away and so the long wavelengths of the error appear 
shorter and can be reduced more efficiently by the same algebraic solver. By using a cascade of progressively coarser grids, all 
wavelengths of the error can be reduced with equal efficiency \cite{Brandt_77}.

The solution procedure is transferred between grids as follows. Let the system of all algebraic equations be written as:

\begin{equation} \label{eq: Nu=b}
 N_h(x_h) \;=\; b_h
\end{equation}
where now the vector $x_h$ stores all the unknowns (velocity components, pressure, and viscosity) at all CV centres of the grid 
whose spacing is $h$. After a few SIMPLE iterations an approximate solution $x_h^*$ is obtained which satisfies the above 
equation up to a residual $r_h$:

\begin{equation} \label{eq: Nh(u)=Nh(u*)-rh}
 N_h(x_h^*) \;=\; b_h \;-\; r_h \;=\; N_h(x_h) \;-\; r_h \Rightarrow \; N_h(x_h) \;=\; N_h(x_h^*) \;+\; r_h
\end{equation}

The algebraic system (\ref{eq: Nh(u)=Nh(u*)-rh}) can be approximated on a coarser grid of spacing $2h$ (grid $2h$ hereafter), 
obtained by removing every second grid line of grid $h$, as follows:

\begin{equation} \label{eq: N2h(u)=N2h(u*)-r2h}
 N_{2h}(x_{2h}) \;=\; N_{2h}(I_h^{2h}x_h^*) \;+\; \hat{I}_h^{2h}r_h
\end{equation}
The operator $N_{2h}$ is constructed on grid $2h$ using the same discretisation schemes as $N_h$ on grid $h$, with $2\Delta x$ 
and $2\Delta y$ used instead of $\Delta x$ and $\Delta y$. The \textit{restriction} operator $I_h^{2h}$ transfers the variables 
from grid $h$ to grid $2h$. The notation comes from the identity matrix $I$, since the variables are not transformed but rather 
transferred from one grid to another. In the present study this restriction operator sets the value of a variable at the centre 
of a control volume $P$ of grid $2h$ equal to the average of the values of that variable over the 4 CVs of grid $h$ that overlap 
with $P$, which will hereafter be called the \textit{children} of $P$. The (possibly different) operator $\hat{I}_h^{2h}$ is used 
for transferring the residuals to grid $2h$. In the present study the residuals at a control volume $P$ of grid $2h$ are set 
equal to the sum of the residuals of all its children. Since the residuals are essentially flux imbalances, this means that the 
flux imbalance of $P$ is set equal to the sum of the imbalances of its children.

The right hand side of (\ref{eq: N2h(u)=N2h(u*)-r2h}) is known, so the system can be solved to obtain $x_{2h}$. Then, the 
\textit{correction} $x_{2h}-I_h^{2h}x_h^*$ is transferred back to grid $h$, to obtain a better estimate $x_h^{new}$ of the exact 
solution $x_h$, and SIMPLE iterations can resume on grid $h$:

\begin{equation} \label{eq: MG correction}
 x_h^{new} \;=\; x_h^* \;+\; I_{2h}^h \left( x_{2h}-I_h^{2h}x_h^* \right)
\end{equation}
The \textit{prolongation} operator $I_{2h}^h$ transfers the correction from grid $2h$ to grid $h$; in the present study linear 
interpolation is used. It is straightforward to see that if in (\ref{eq: Nh(u)=Nh(u*)-rh}) the exact solution has already been 
obtained ($x_h^*=x_h$) then the residuals will be zero and therefore so will be the last term of (\ref{eq: N2h(u)=N2h(u*)-r2h}). 
Then the solution of (\ref{eq: N2h(u)=N2h(u*)-r2h}) will be $x_{2h}=I_h^{2h}x_h^*$ (it is important that $I_h^{2h}x_h^*$ be used 
as the initial guess when solving (\ref{eq: N2h(u)=N2h(u*)-r2h})), and therefore the correction $x_{2h}-I_h^{2h}x_h^*$ will be 
zero. So, (\ref{eq: MG correction}) will leave the solution unaltered: $x_h^{new} = x_h^* = x_h$.

\begin{figure}[htb]
 \centering
 \noindent\makebox[\textwidth]{
 \includegraphics[scale=0.75]{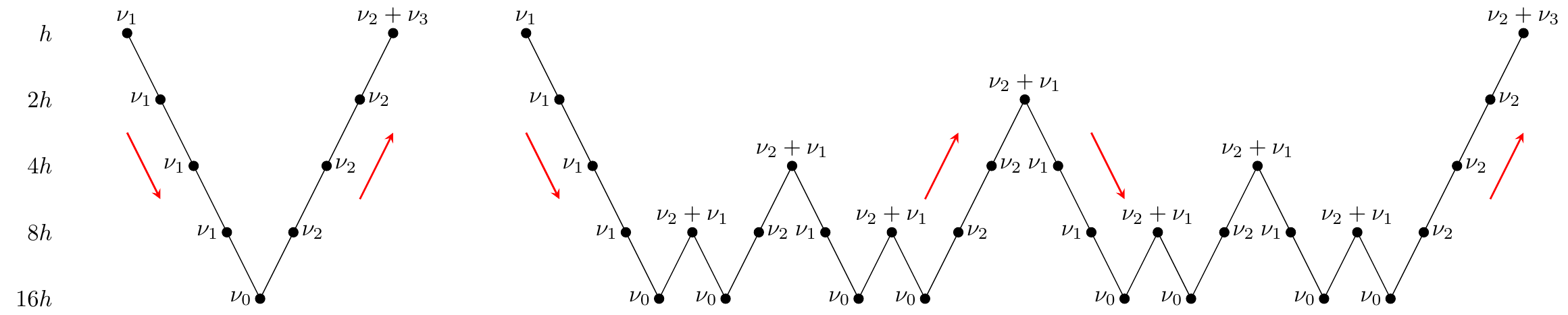}}
 \caption{Multigrid cycles: V($\nu_1,\nu_2$)--$\nu_3$ (left) and W($\nu_1,\nu_2$)--$\nu_3$ (right) cycles, using 5 grid levels.}
 \label{fig: multigrid cycles}
\end{figure}

The coarse grid problem (\ref{eq: N2h(u)=N2h(u*)-r2h}) can be solved using an even coarser grid $4h$ and so on. This whole 
process of going down from grid $h$ to some coarsest grid and then back again up to grid $h$ constitutes a \textit{multigrid 
cycle}. In general, a number of multigrid cycles will be required to reduce the residuals by a given amount. Different kinds of 
cycles have been suggested and used. The most popular are the V($\nu_1,\nu_2$) and W($\nu_1,\nu_2$) cycles. In V($\nu_1,\nu_2$) 
cycles, after performing $\nu_1$ iterations on the fine grid $h$, the coarse grid system (\ref{eq: N2h(u)=N2h(u*)-r2h}) is solved 
using one multigrid cycle, and then the coarse grid correction is transferred back to grid $h$ according to (\ref{eq: MG 
correction}), where another $\nu_2$ iterations are performed. The difference between the W and V cycles is that in W cycles the 
coarse grid problems such as (\ref{eq: N2h(u)=N2h(u*)-r2h}) are solved using two rather than one cycle. In the present work we 
have observed that it is sometimes useful to perform $\nu_3$ extra SIMPLE iterations on the finest grid $h$ only, between cycles. 
Such cycles will be denoted V($\nu_1,\nu_2$)--$\nu_3$ or W($\nu_1,\nu_2$)--$\nu_3$ respectively. These cycles are shown 
schematically in Figure \ref{fig: multigrid cycles}. The costs of a V($\nu_1,\nu_2$)--$\nu_3$ and a W($\nu_1,\nu_2$)--$\nu_3$ 
cycle are approximately equal to $[\frac{4}{3}(\nu_1+\nu_2)+\nu_3]$ and $[2(\nu_1+\nu_2)+\nu_3]$ times the cost of a single 
iteration on the finest grid $h$, respectively.

\section{Numerical results}
\label{sec: results}

The creeping flow ($Re=0$) of a Bingham plastic in a square lid-driven cavity has been chosen as the test case for the finite 
volume method described in the previous section. The computational domain is a square, enclosed by solid boundaries of equal 
length. The top boundary (lid) moves towards the right with a uniform horizontal velocity, while the rest of the boundaries are 
still. The problem was solved for a range of Bingham numbers, up to $1000$. In all cases an exponent $M=400$ was used, except for 
the two highest Bingham numbers tested, $Bn=500$ and $1000$, for which $M=100$ was used due to convergence difficulties. In order 
to check grid convergence, the domain was discretised using three uniform Cartesian grids with $64 \times 64$, $128 \times 128$ 
and $256 \times 256$ CVs. Unless otherwise stated, the results presented below were calculated on the $256 \times 256$ grid. The 
pressure was set to zero at the centre of the domain.

\begin{figure}[p]
% \begin{figure}[!htb]
\centering
\noindent\makebox[\textwidth]{
 \subfigure[{$Bn = 2$}] {\label{sfig: streamlines Bn=2}
  \includegraphics[scale=0.09]{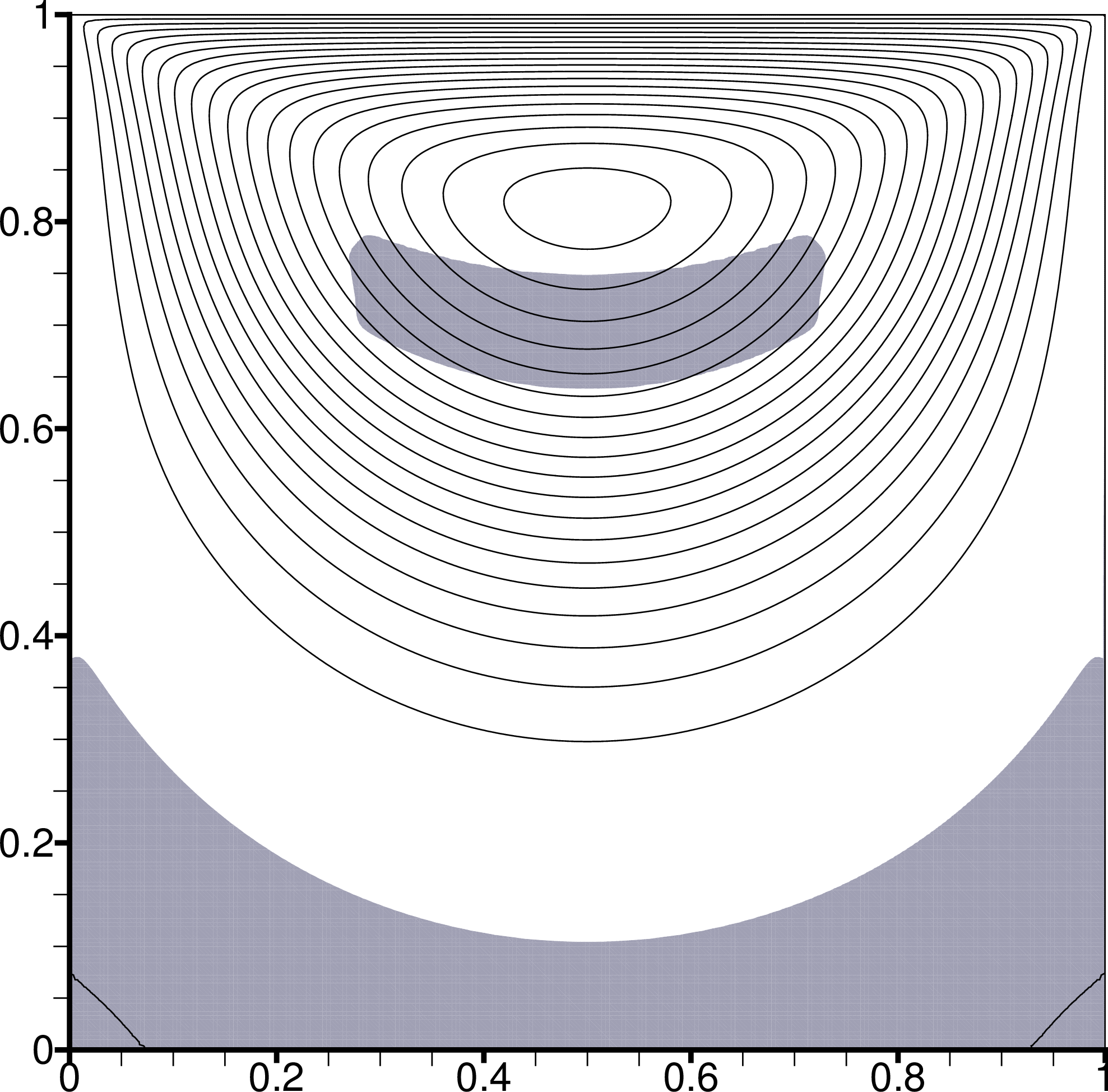}}
 \subfigure[{$Bn = 5$}] {\label{sfig: streamlines Bn=5}
  \includegraphics[scale=0.09]{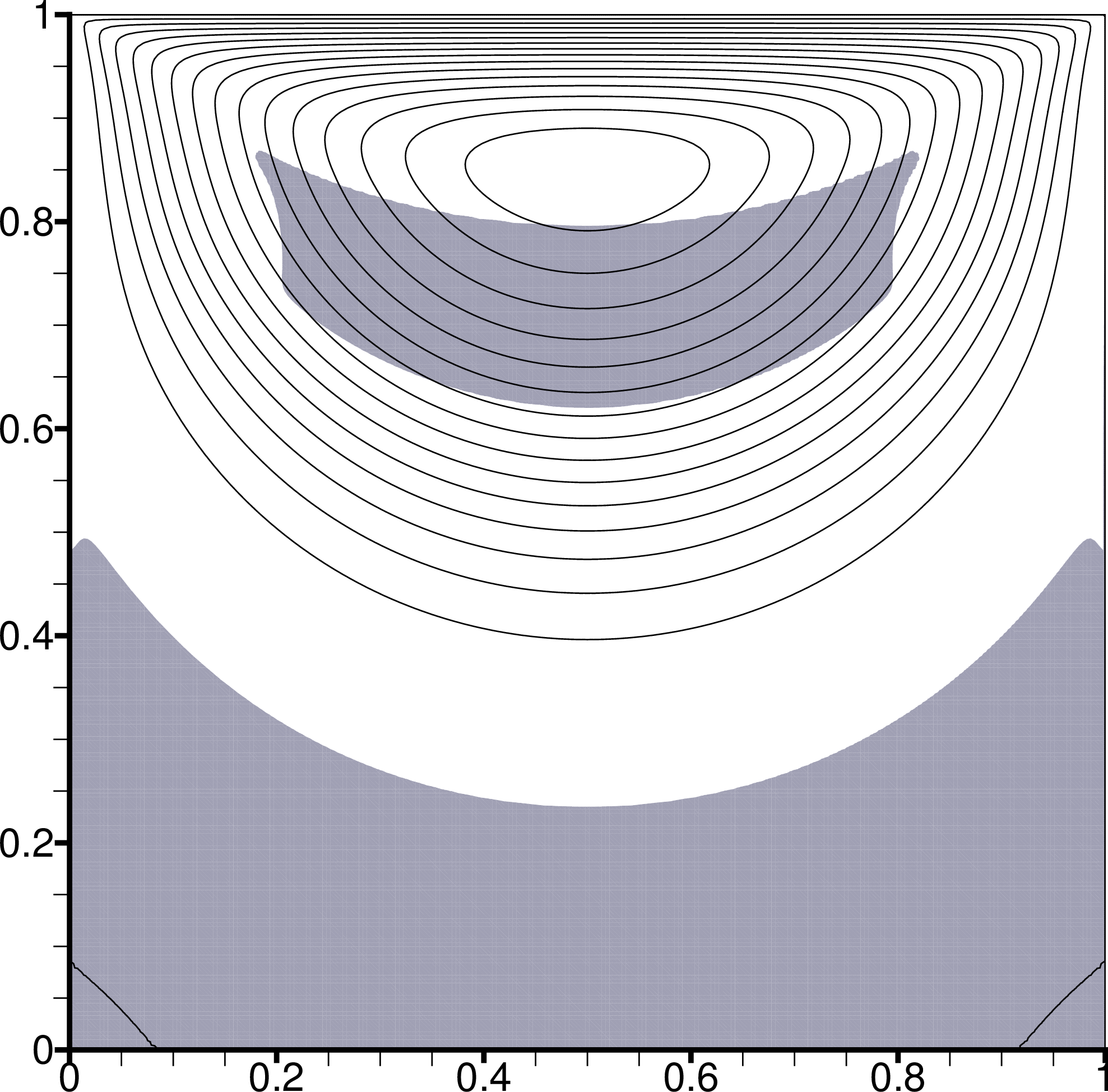}}
}
\noindent\makebox[\textwidth]{
 \subfigure[{$Bn = 20$}] {\label{sfig: streamlines Bn=20}
  \includegraphics[scale=0.09]{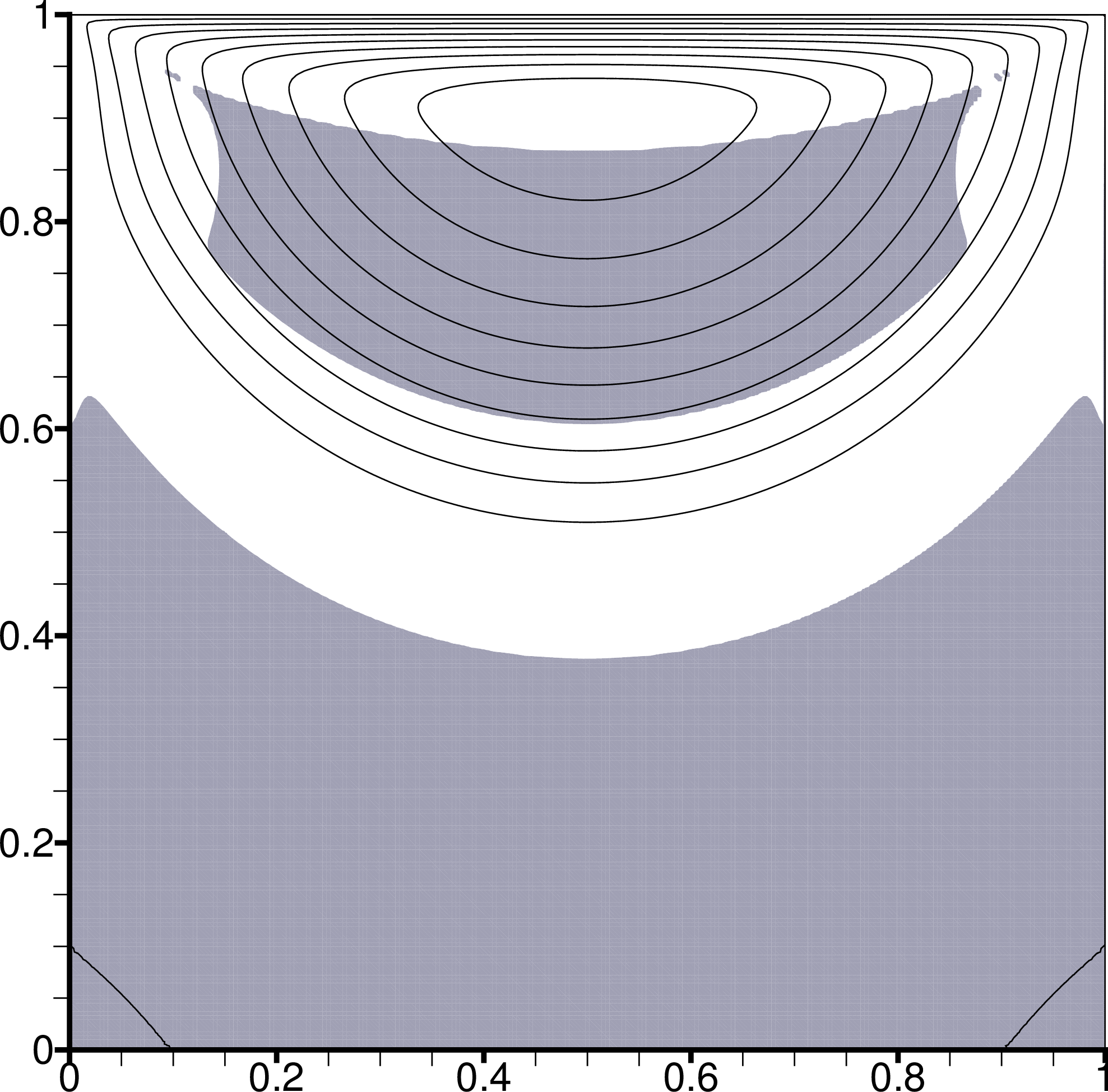}}
 \subfigure[{$Bn = 50$}] {\label{sfig: streamlines Bn=50}
  \includegraphics[scale=0.09]{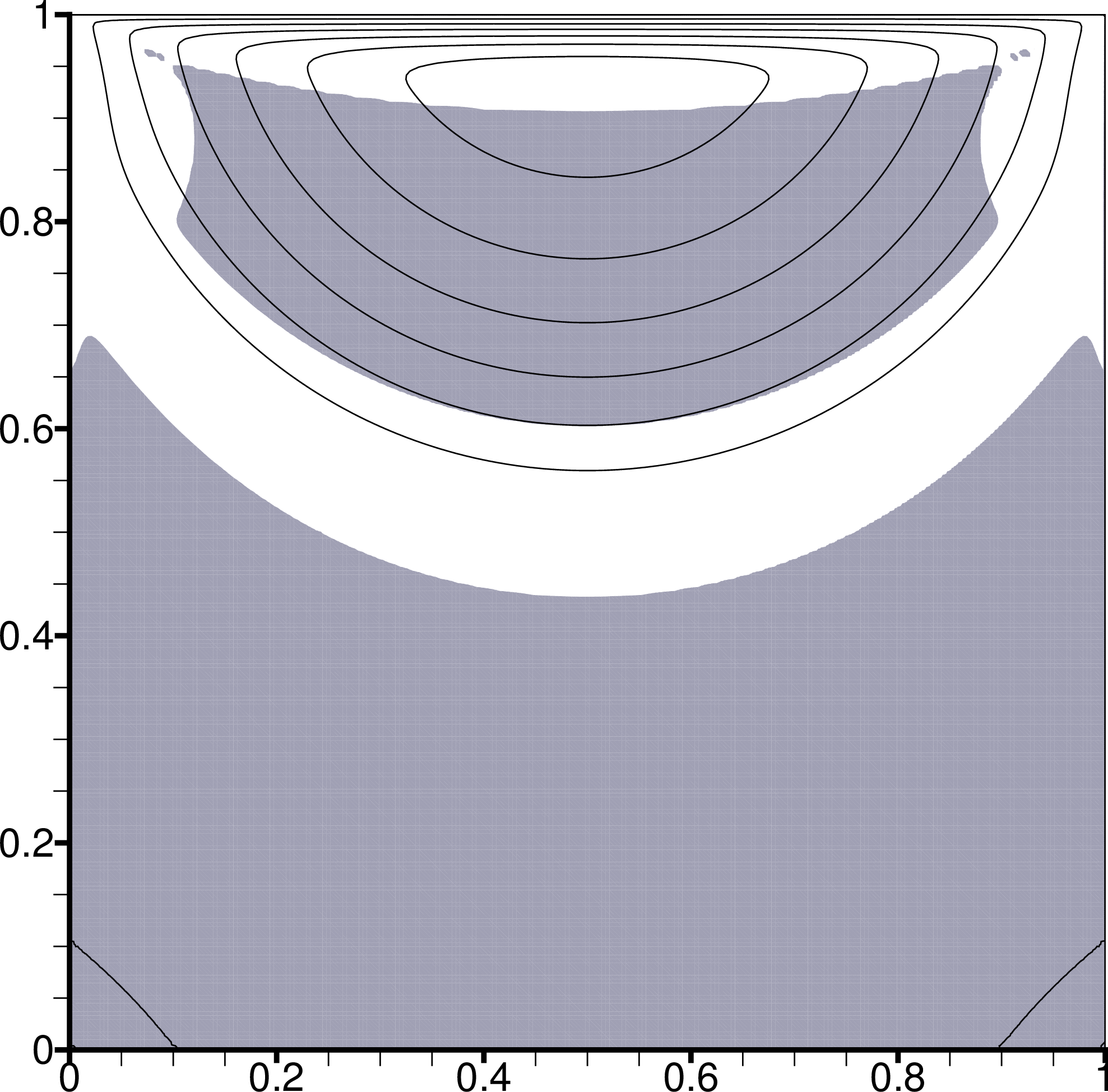}}
}
\noindent\makebox[\textwidth]{
 \subfigure[{$Bn = 200$}] {\label{sfig: streamlines Bn=200}
  \includegraphics[scale=0.09]{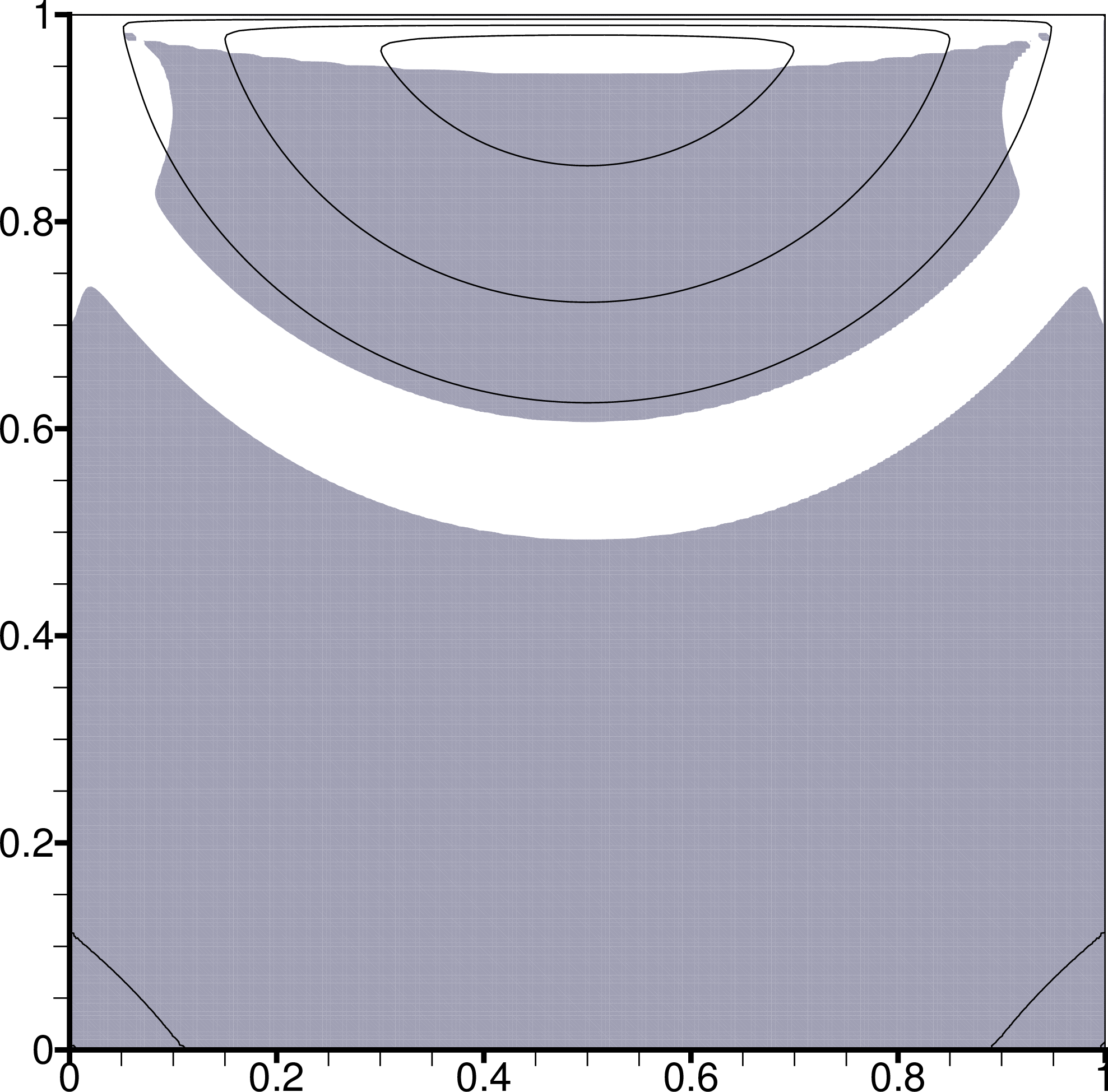}}
 \subfigure[{$Bn = 500$}] {\label{sfig: streamlines Bn=500}
  \includegraphics[scale=0.09]{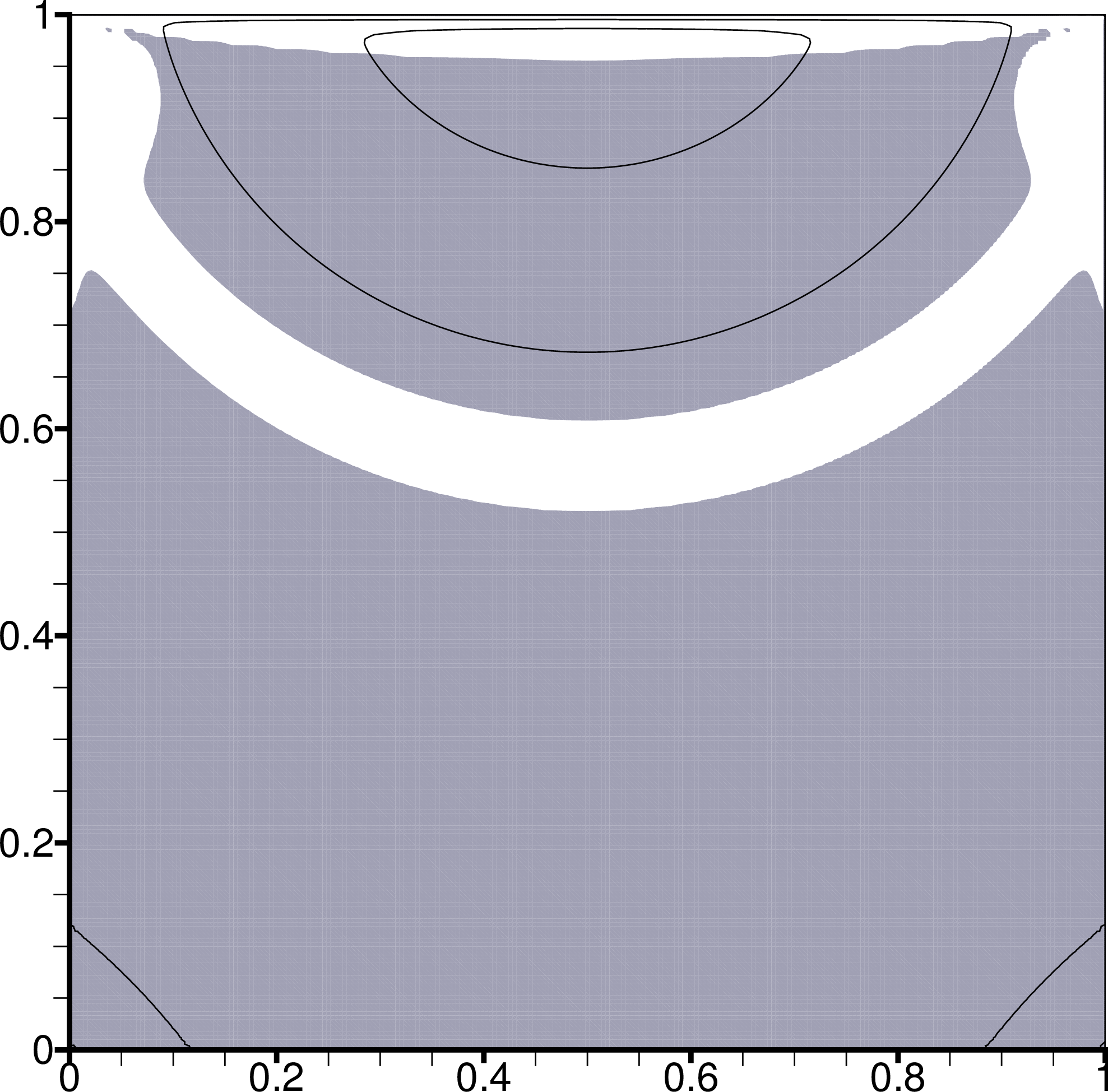}}
}
\caption{Streamlines plotted at intervals of 0.004 starting from zero. Unyielded areas ($\tau < Bn$) are shown shaded.}
\label{fig: streamlines}
\end{figure}

Figure \ref{fig: streamlines} shows the streamlines calculated for selected Bingham numbers. These are the Bingham numbers chosen 
also by Mitsoulis and Zisis \cite{mitsoulis_01} and Yu and Wachs \cite{Yu_07} for their corresponding figures. The ``unyielded''  
regions are also shown. These are defined here as the regions where $\tau < Bn$, where $\tau = \eta \dot{\gamma}$ is calculated 
from (\ref{eq: papanastasiou_eta nd}) and (\ref{eq: gamma discretised}). Of course, they are not actually unyielded, but they 
approximate the unyielded regions of an ideal Bingham flow. The results generally agree with those presented by other 
researchers, in \cite{mitsoulis_01, Yu_07, Olshanskii_09, Zhang_10}. The agreement is good with the results of Mitsoulis and 
Zisis \cite{mitsoulis_01} in the whole range of Bingham numbers, except that these authors predict yielded regions with more 
rounded corners than the present results. Also, their yield surfaces are not smooth, but this may be due to the low resolution of 
the grid that they employed. The results also agree well with those of Yu and Wachs \cite{Yu_07} up to a Bingham number of 20, 
but there are notable differences at higher Bingham numbers - however, Yu and Wachs state that, as far as capturing the yield 
surface is concerned, their method performs well at low to moderate Bingham numbers but less so at high Bingham numbers. Direct 
comparison can also be made between the present results and those of Olshanskii \cite{Olshanskii_09} for $Bn=2$ and $Bn=5$, who 
used an augmented Lagrangian approach instead of a regularisation method, and the agreement is very good. Zhang \cite{Zhang_10}, 
who also used an augmented Lagrangian approach, also provides results for $Bn=2,20,50$ and $200$, where the unyielded regions 
appear somewhat smaller, flatter, and rounded compared to the present results, although qualitatively similar.

The main characteristic of the flow field is the vortex which develops at the upper central region of the cavity, similarly to 
the Newtonian case. Two distinct unyielded regions can be observed: A larger one at the bottom of the cavity, and a smaller one 
just below the vortex centre. The flow field is symmetric with respect to the vertical centreline. As the $Bn$ number increases, 
the unyielded regions expand and the flow circulation becomes weaker and limited to the upper part of the cavity, with the centre 
of the vortex coming closer to the lid. One can notice in Figure \ref{fig: streamlines} that a couple of secondary vortices 
appear at the lower two corners of the cavity, which are completely inside the lower unyielded region. These vortices, which 
appear to slightly grow in size as $Bn$ increases, are an artifact of the regularisation of the Bingham model by the 
Papanastasiou approximation. In fact, the velocity is extremely low throughout the lower unyielded region. In reality, since this 
region is in contact with the rigid walls which are motionless, and a no-slip boundary condition applies, the unyielded material 
should also be motionless throughout.

The situation is different at the upper unyielded region, where the velocity is non-zero as can be clearly seen from the 
streamlines' spacing. So, the upper unyielded regions move as solid bodies. In the present case it appears that the material at 
the upper unyielded region moves as a solid body, and in particular it rotates, with the streamlines forming circular arcs inside 
it. The fact that the streamlines cross into this region means that the fluid ``solidifies'' on entry, and becomes fluid again on 
exit from the region.

\begin{figure}[p]
% \begin{figure}[!htb]
\centering
\subfigure[{Vortex centre}] {\label{sfig: vortex centre}
  \includegraphics[scale=0.56]{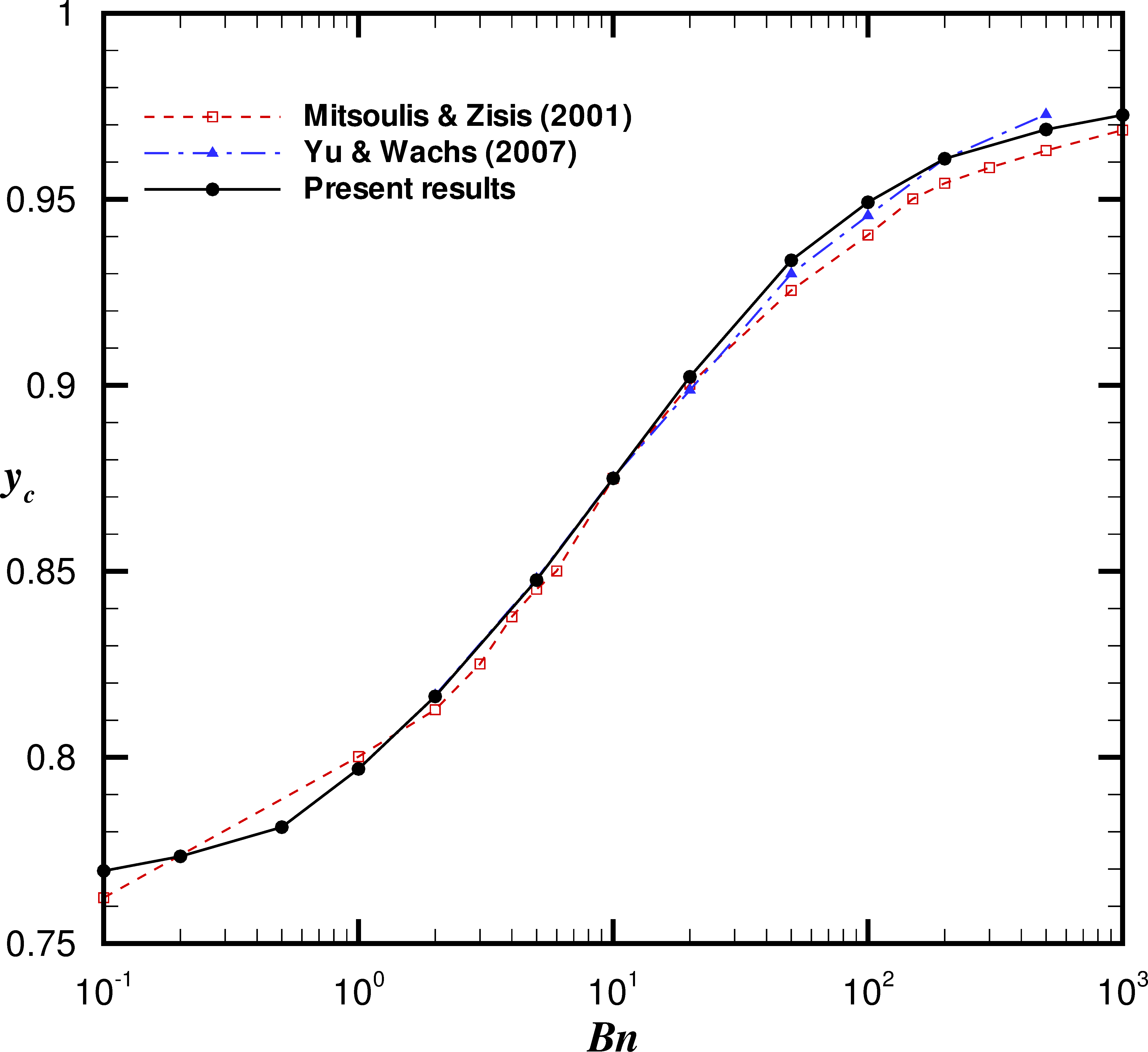}}
\subfigure[{Vortex strength}] {\label{sfig: vortex strength}
  \includegraphics[scale=0.56]{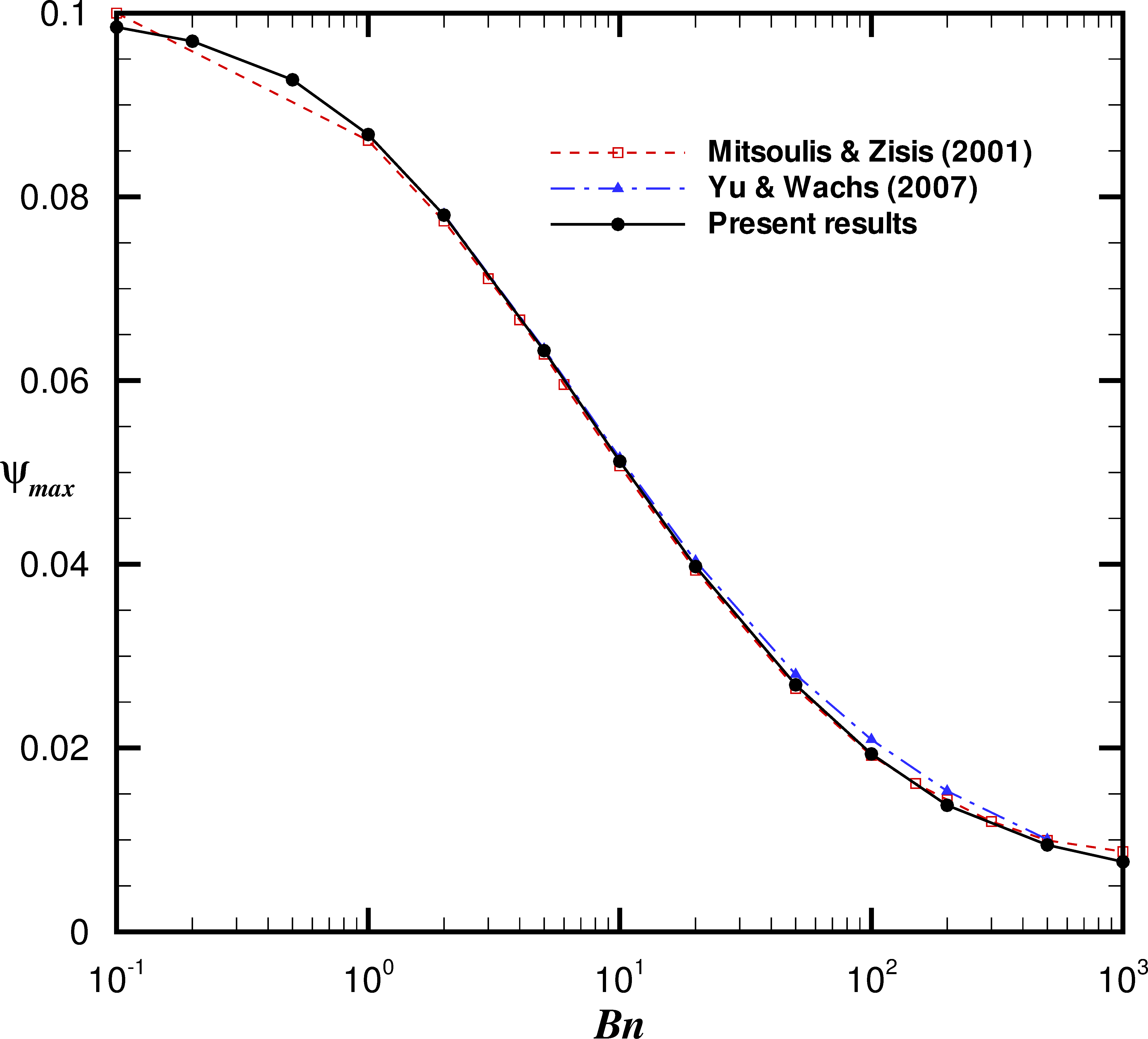}}
\caption{The $y$ coordinate of the centre of the vortex, and vortex strength as a function of the Bingham number. }
\label{fig: main vortex}
\end{figure}

The behaviour of the vortex as a function of the $Bn$ number is described in more detail in Figure \ref{fig: main vortex}, where 
the vertical position and the strength of the vortex are plotted as functions of $Bn$. The weakening of the circulation as $Bn$ 
increases is accompanied by an increase in the stress and pressure levels in the cavity, as greater stresses are needed in order 
to make the material flow. Figure \ref{fig: main vortex} also serves to validate the results of the present study against those 
of other researchers. It can be seen that the results are quite close to previously published data of \cite{mitsoulis_01} and 
\cite{Yu_07}. The results are closer to those of Mitsoulis and Zisis \cite{mitsoulis_01} as far as the vortex strength is 
concerned, and to the results of Yu and Wachs \cite{Yu_07} as far as the vortex position is concerned (possibly because the grid 
of \cite{mitsoulis_01} is coarse).

\begin{figure}[!htb]
\centering
%\noindent\makebox[\textwidth]{
  \subfigure[{$p$ at the lid}] {\label{sfig: p at lid}
    \includegraphics[scale=0.45]{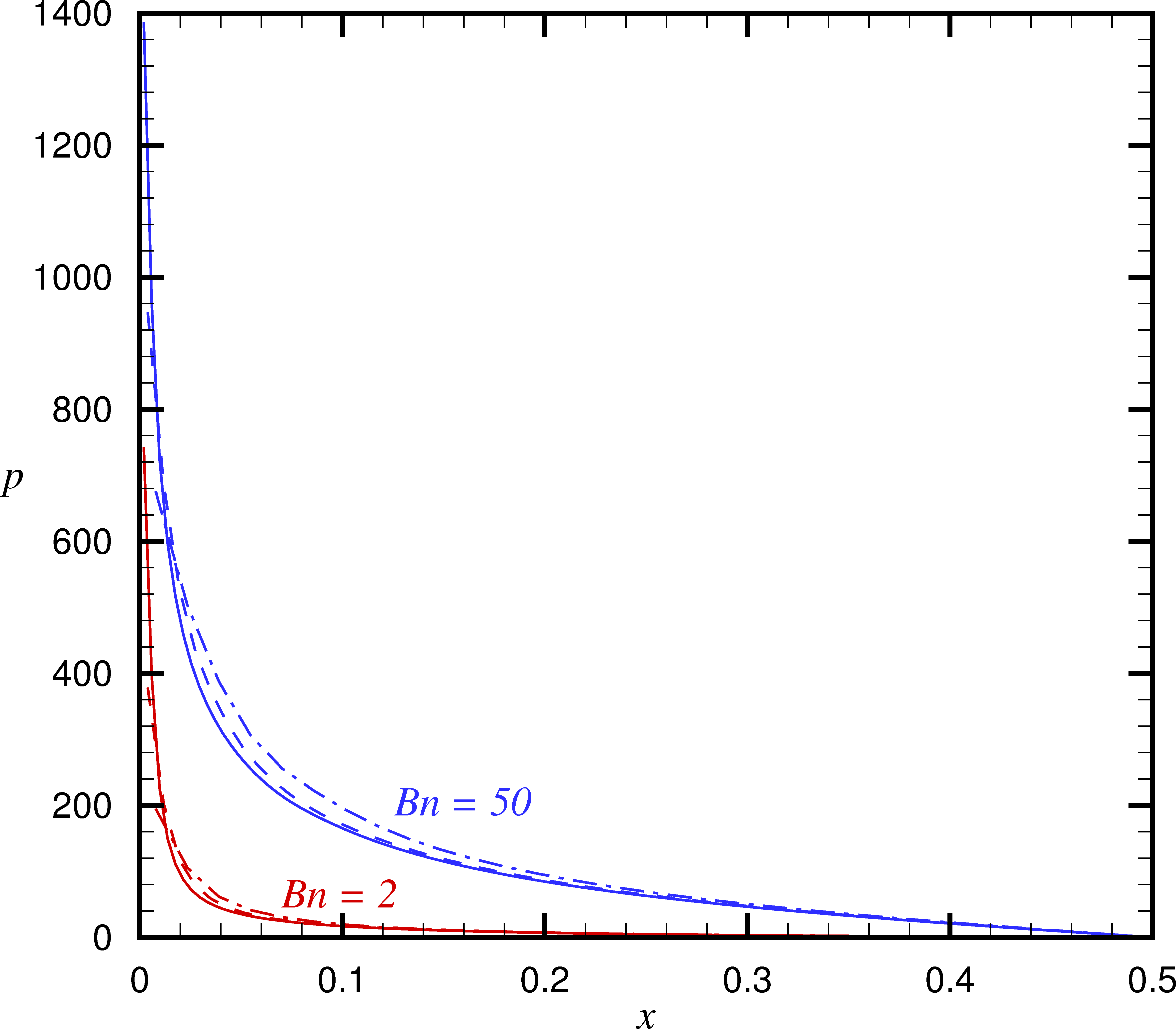}}
  \subfigure[{$\tau_{yx}$ at the lid}] {\label{sfig: tau at lid}
    \includegraphics[scale=0.45]{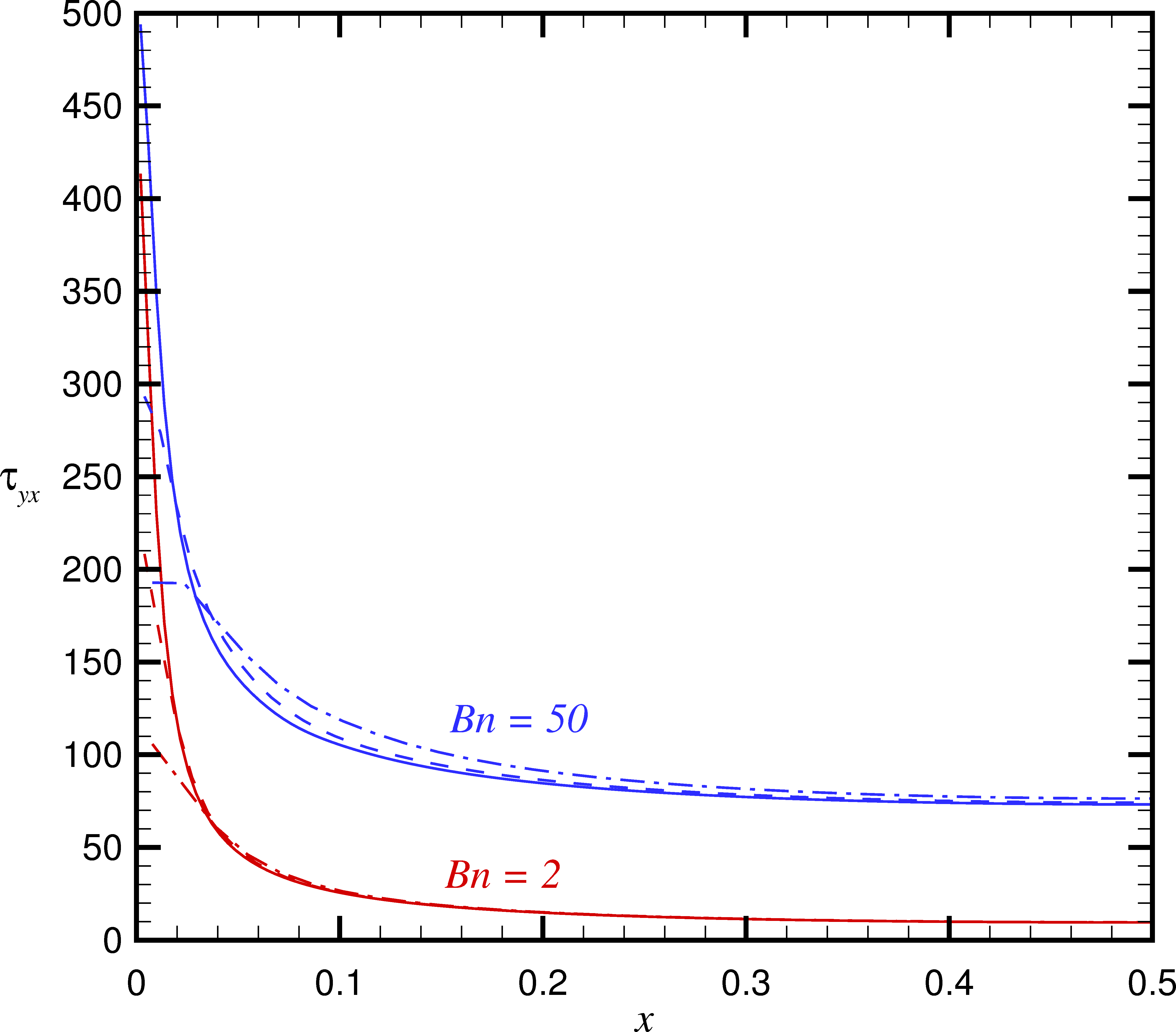}}
%}
\caption{Pressure and shear stress at the left half of the lid for $Bn = 2$ and $50$. Results are shown for grids $64 \times 64$ 
(chained lines), $128 \times 128$ (dashed lines), and $256 \times 256$ (solid lines).}
\label{fig: tau & p at lid}
\end{figure}

\begin{figure}[!hb]
\centering
\noindent\makebox[\textwidth]{
 \subfigure[{$Bn = 0$}] {\label{sfig: pressure Bn=0}
  \includegraphics[scale=0.31]{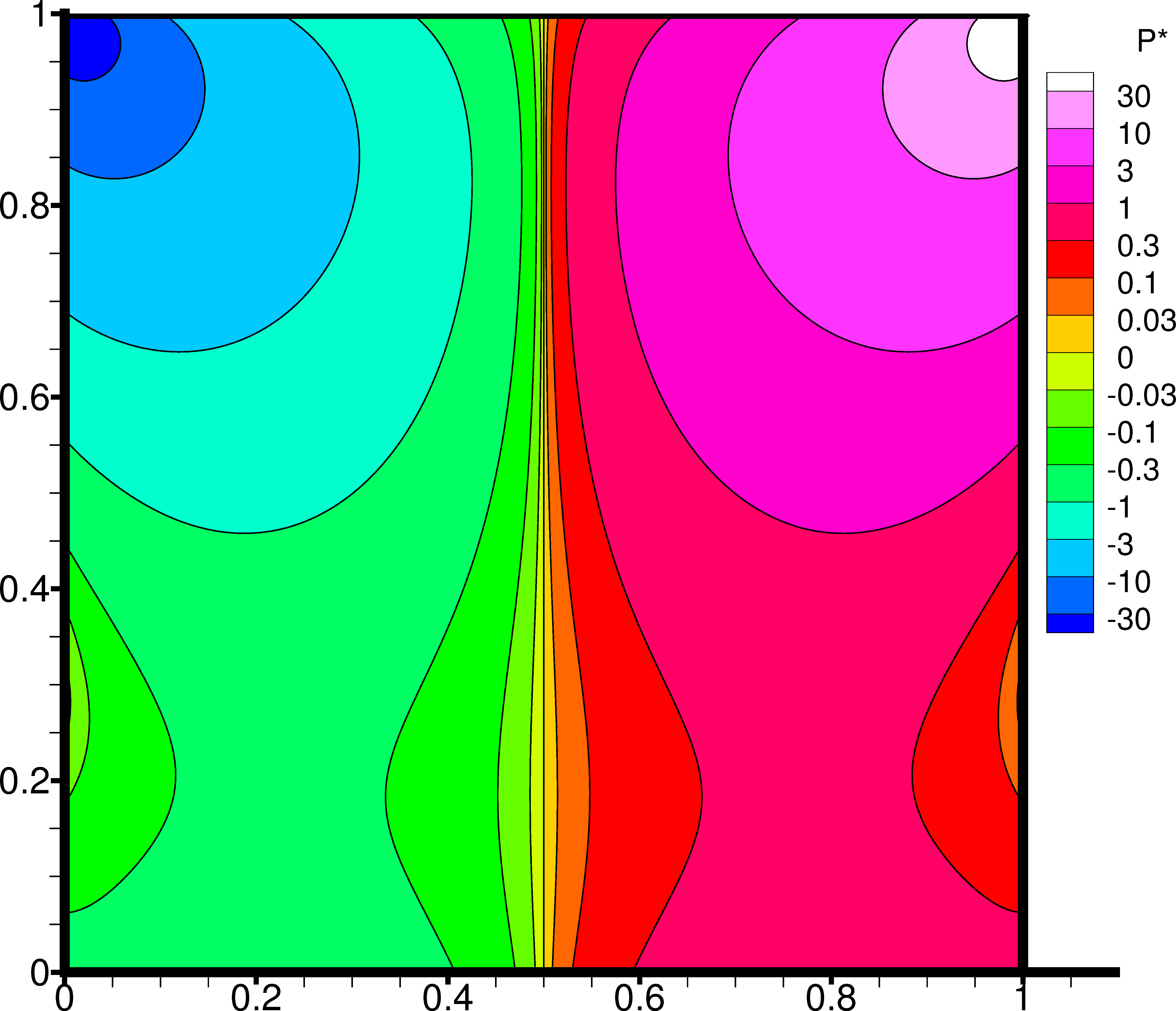}}
 \subfigure[{$Bn = 2$}] {\label{sfig: pressure Bn=2}
  \includegraphics[scale=0.31]{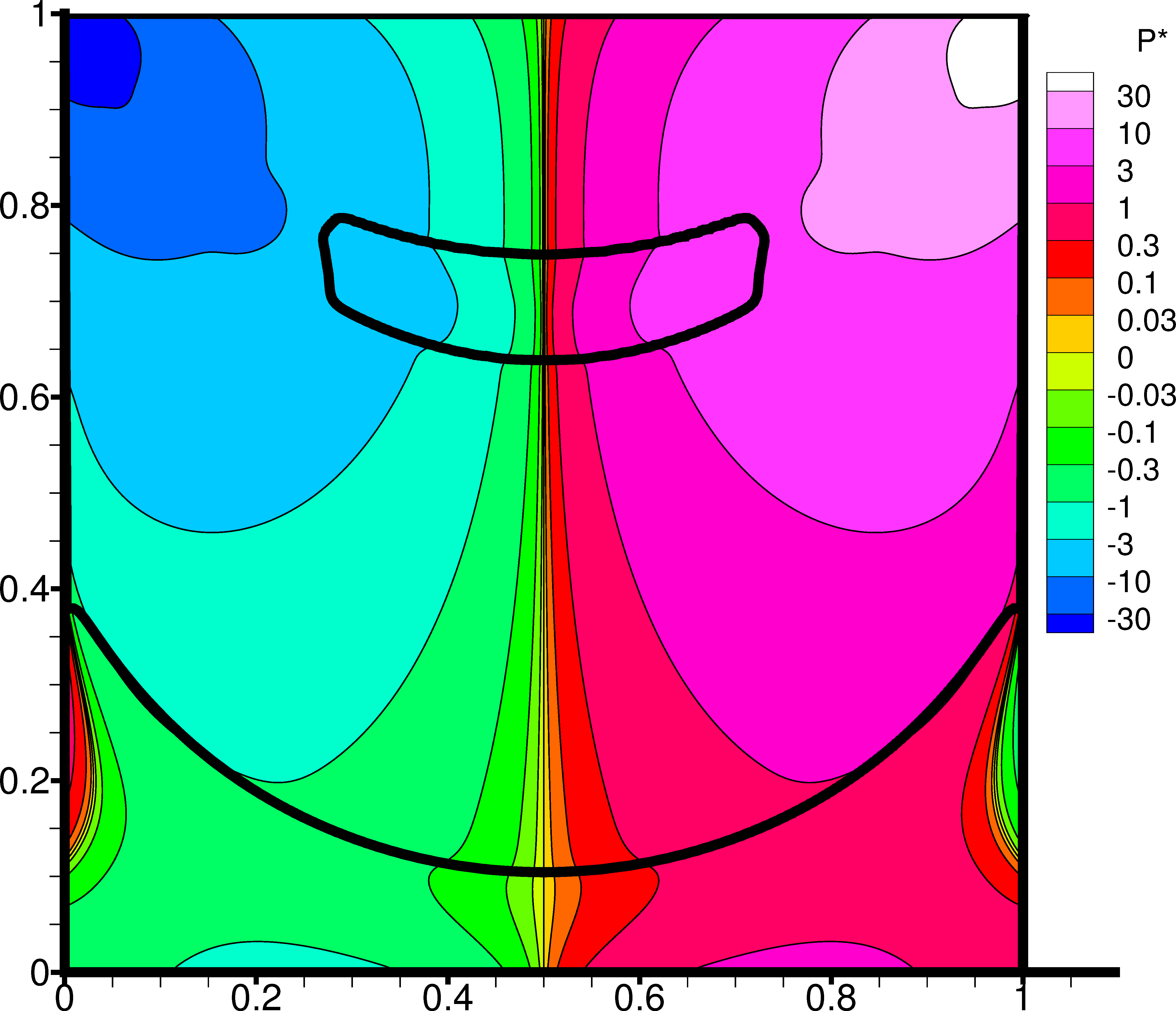}}
 \subfigure[{$Bn = 20$}] {\label{sfig: pressure Bn=20}
  \includegraphics[scale=0.31]{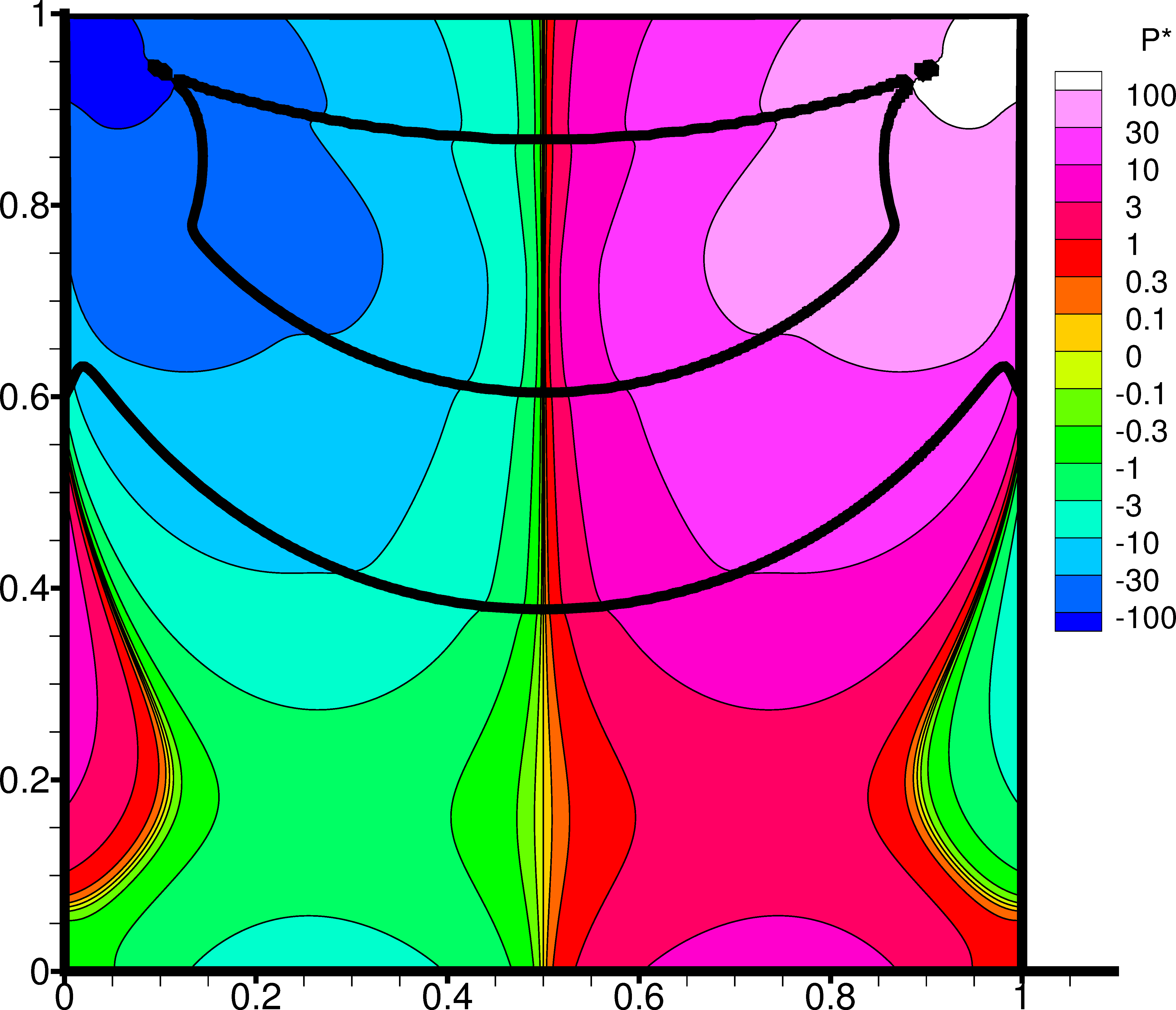}}
}
\caption{Pressure contours for various $Bn$ numbers. For $Bn \neq 0$ the thick lines outline the unyielded regions. Note that the 
pressure scale of the $Bn = 20$ case is different.}
\label{fig: pressure}
\end{figure}

Figure \ref{fig: tau & p at lid} shows the pressure and shear stress distributions at the left half of the lid. Increasing the 
Bingham number from $2$ to $50$ causes a significant increase in pressure and stress, as discussed previously. Figure \ref{fig: 
tau & p at lid} can also be used to verify grid convergence, as it contains results on the three grids used in this work. In the 
interior of the lid the pressure and stress converge with grid refinement, and the rate of convergence is that expected of a 
2$^\text{nd}$ order method. However, grid refinement causes the pressure and stress to tend to infinity at the lid corners. This 
is because of the discontinuity which exists at these corners: the velocity jumps from zero (side walls) to one (lid). Due to 
this singularity, it was observed that the pressure and stress integrals over the lid do not converge with grid refinement, and 
so the shear force needed to drive the lid cannot be calculated accurately. The results on grid convergence are discussed later 
on. The increase in pressure with the Bingham number can also be seen in the pressure contours of Figure \ref{fig: pressure}. It 
is clear that the presence of the unyielded regions causes a distortion in the pressure field. The streamwise pressure gradients 
are greater inside the upper unyielded region.

\begin{figure}[!p]
\centering
\noindent\makebox[\textwidth]{
 \subfigure[{$Bn = 2$}] {\label{sfig: yield lines Bn=2}
  \includegraphics[scale=0.5]{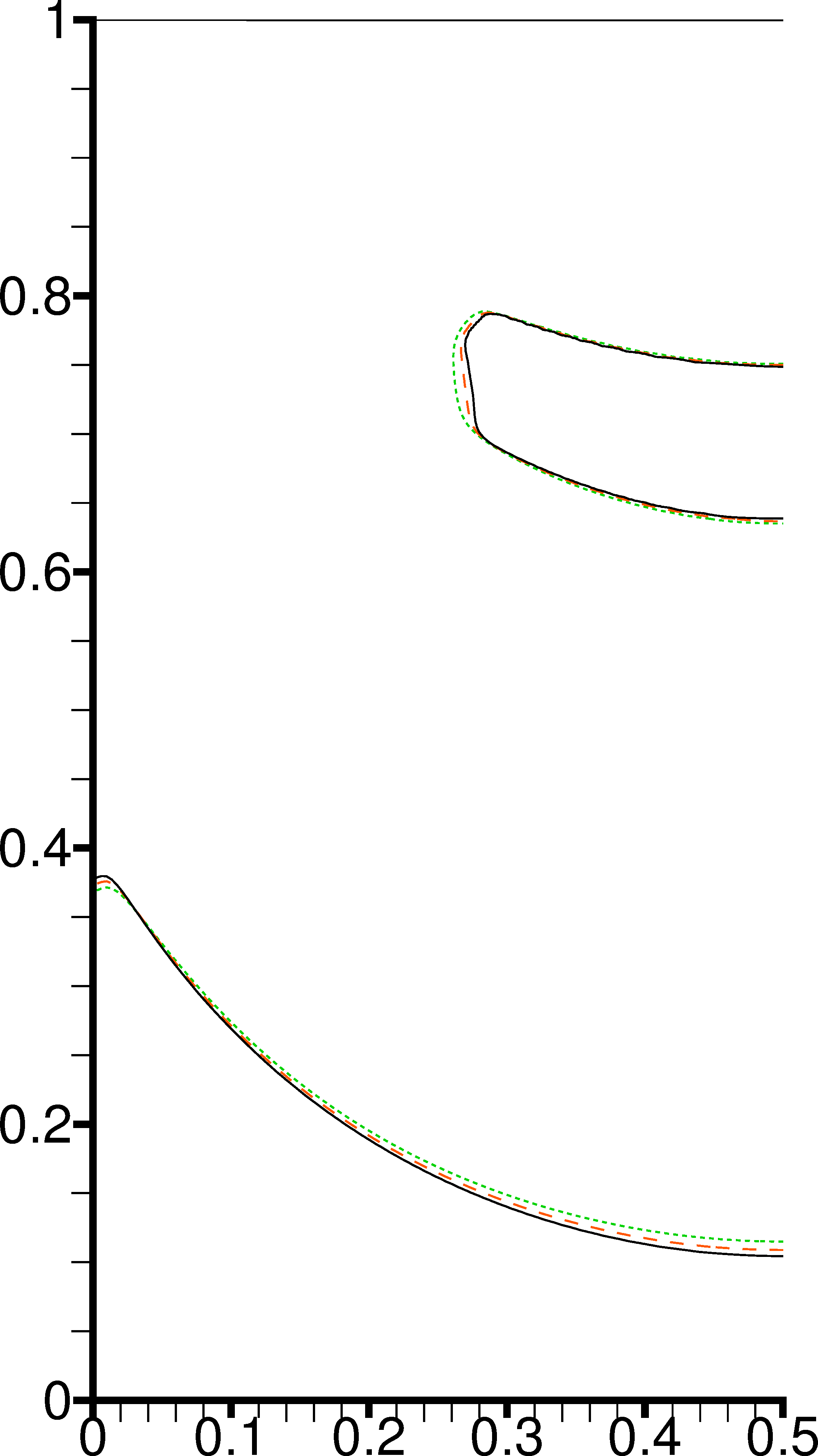}}
 \subfigure[{$Bn = 20$}] {\label{sfig: yield lines Bn=20}
  \includegraphics[scale=0.5]{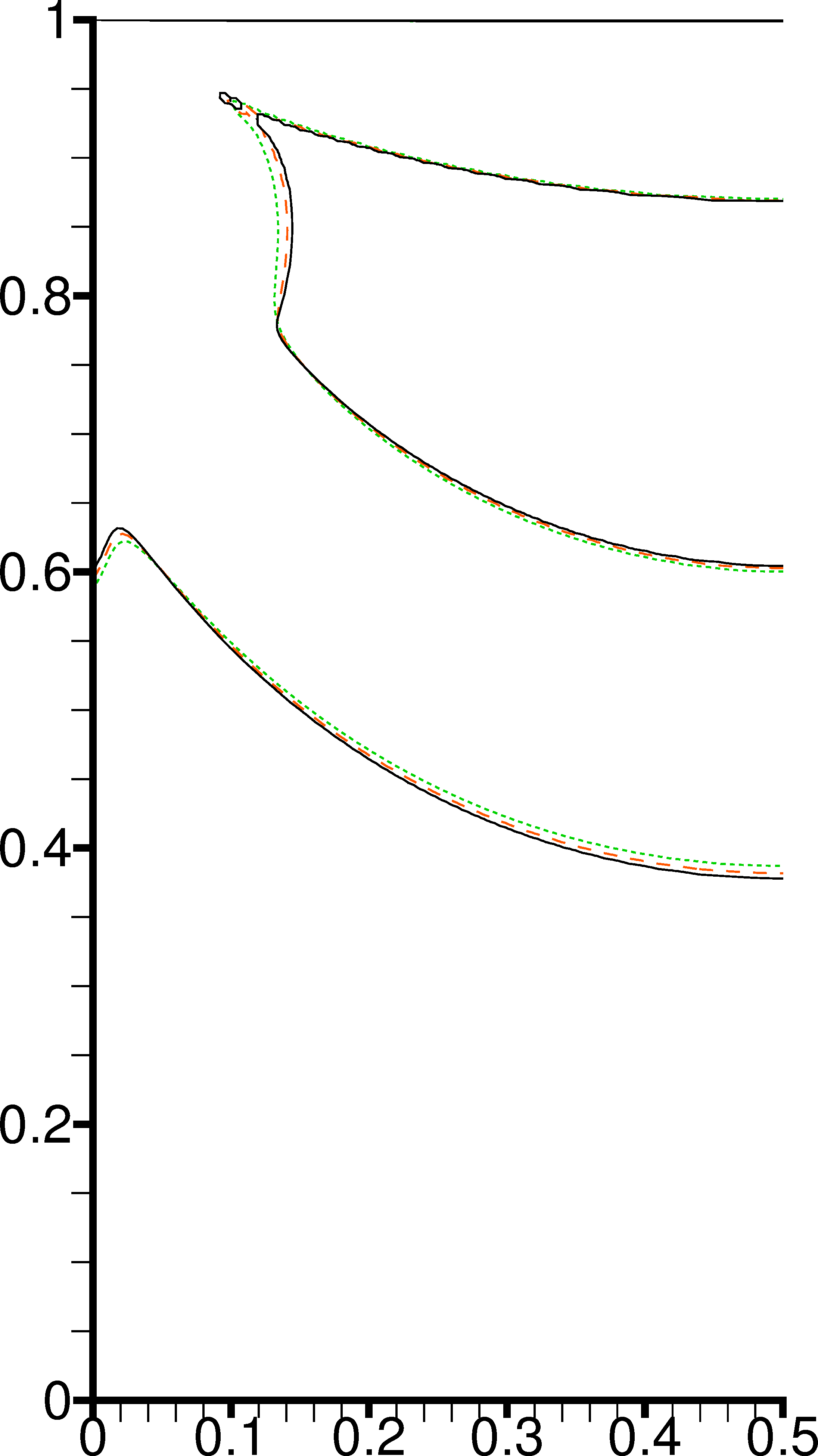}}
 \subfigure[{$Bn = 200$}] {\label{sfig: yield lines Bn=200}
  \includegraphics[scale=0.5]{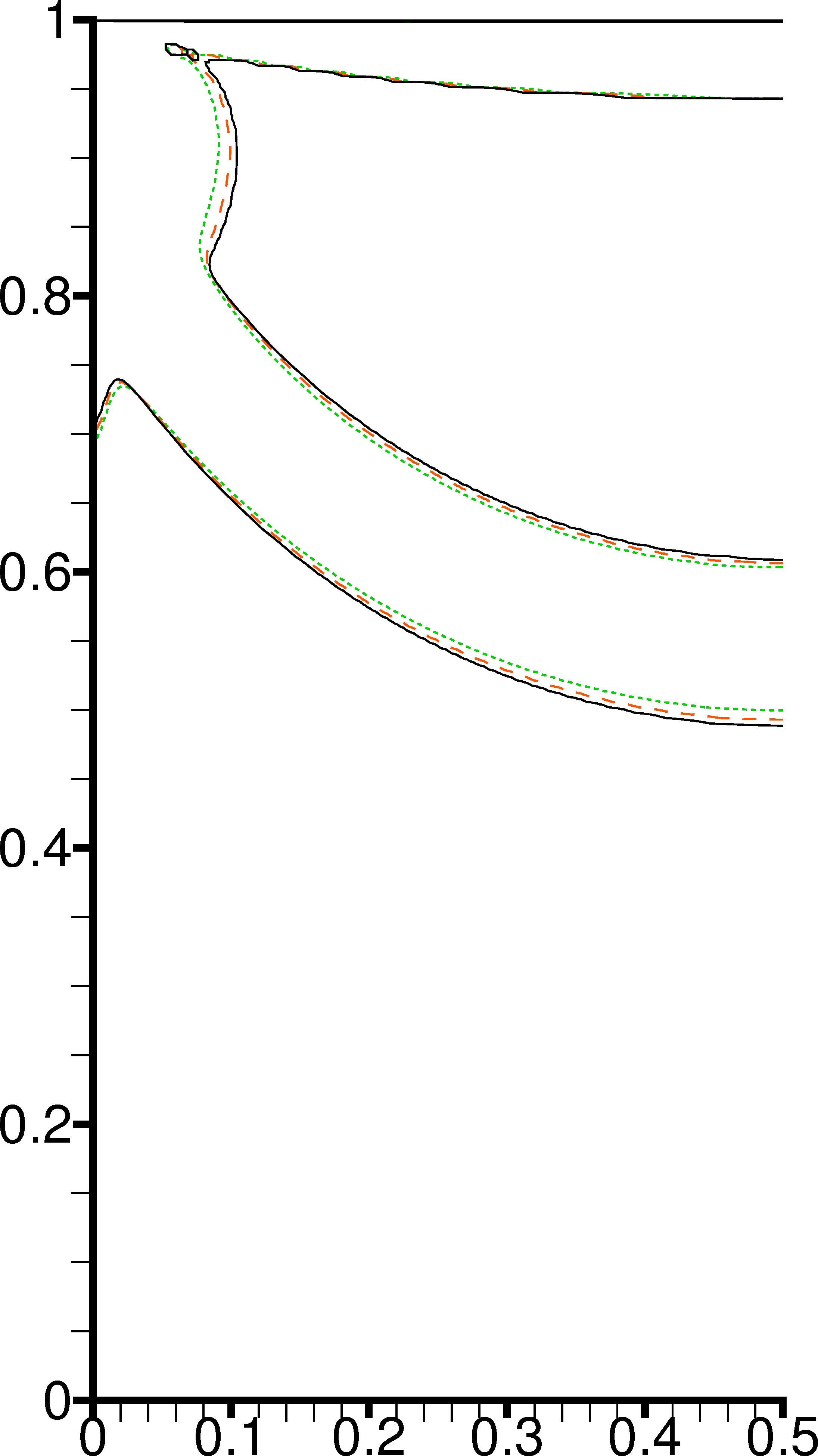}}
}
 \caption{Yield lines ($\tau = Bn$) calculated with $M=100$ (dotted), $M=200$ (dashed) and $M=400$ (solid).}
\label{fig: yield lines}
\end{figure}

\begin{figure}[!p]
\centering
\noindent\makebox[\textwidth]{
 \subfigure[{$Bn = 2$, $\tau = (1+0.01)Bn$}] {\label{sfig: tau=1.01ty Bn=2}
  \includegraphics[scale=0.5]{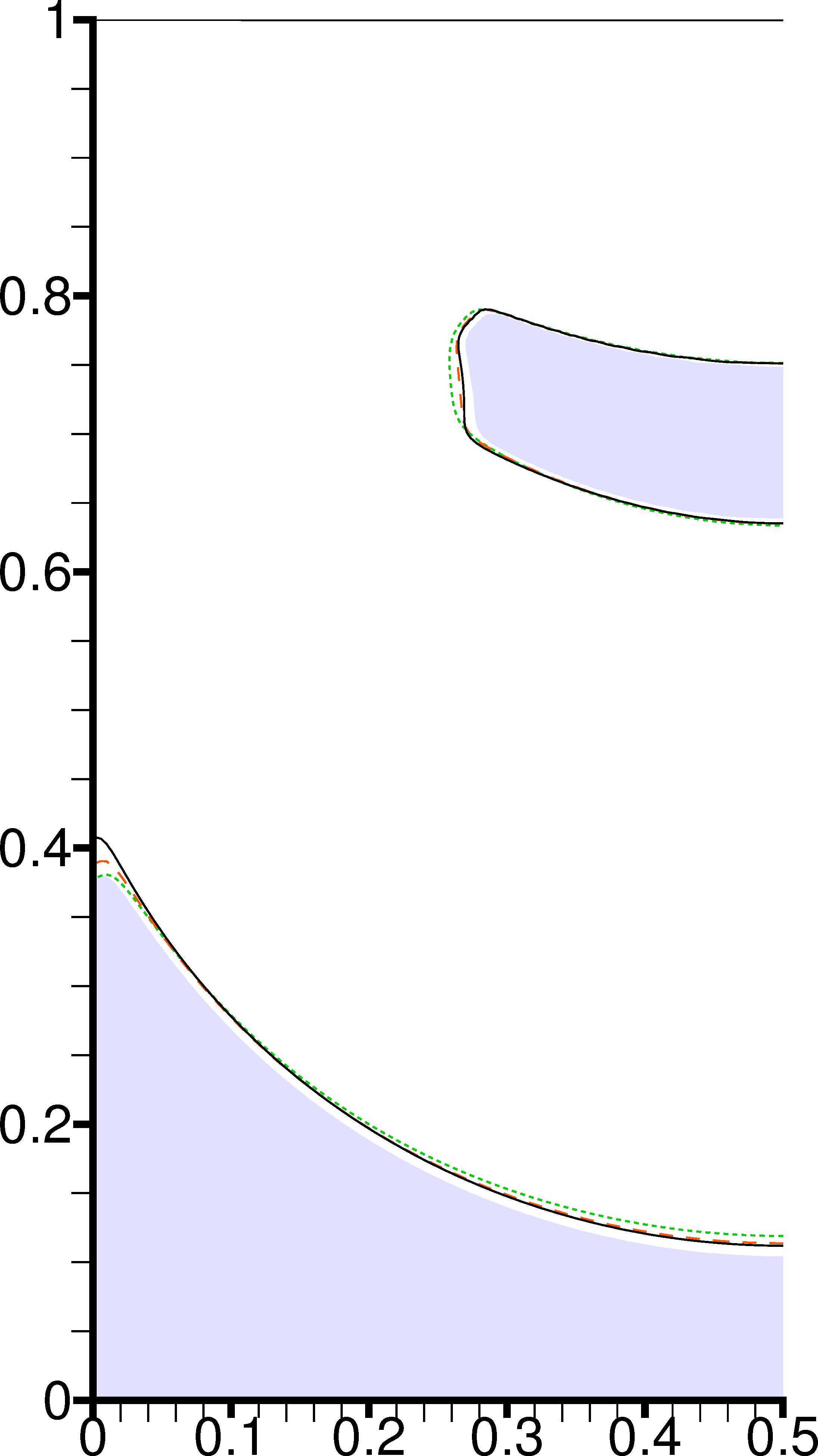}}
 \subfigure[{$Bn = 20$, $\tau = (1\pm 0.01)Bn$}] {\label{sfig: tau=1.01ty Bn=20}
  \includegraphics[scale=0.5]{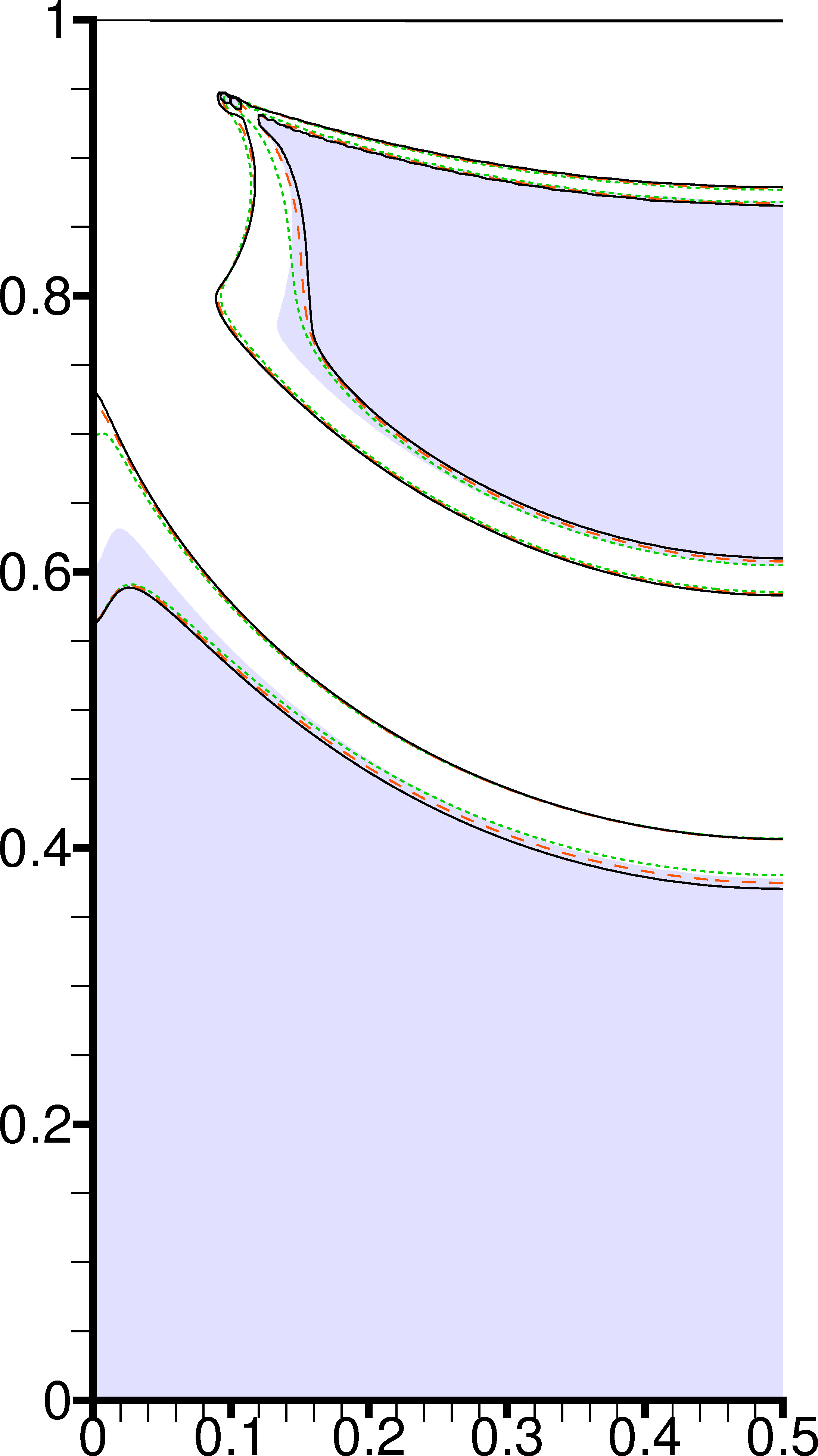}}
 \subfigure[{$Bn = 200$, $\tau = (1\pm 0.01)Bn$}] {\label{sfig: tau=1.01ty Bn=200}
  \includegraphics[scale=0.5]{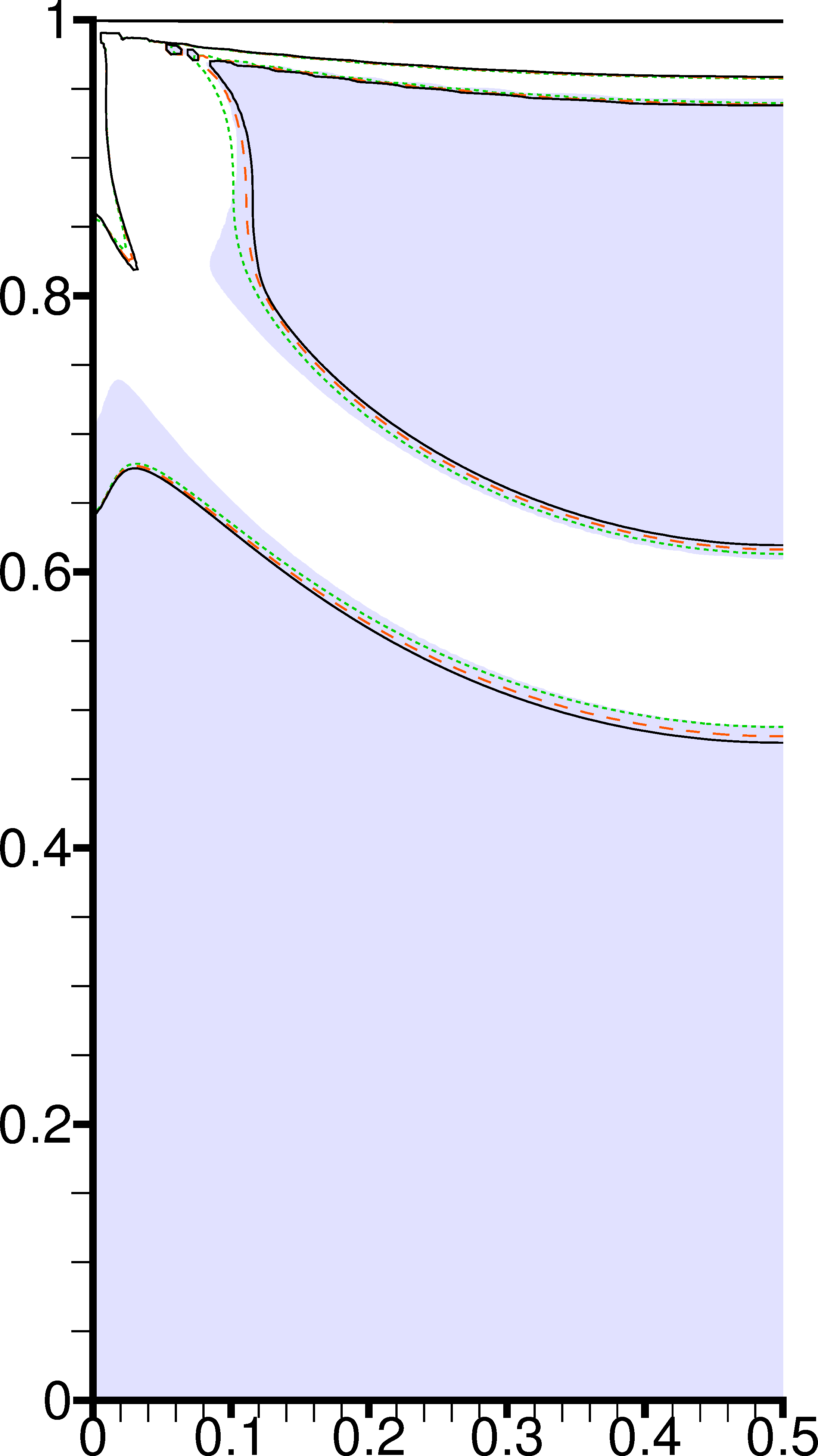}}
}
 \caption{Iso-stress lines calculated with $M=100$ (dotted), $M=200$ (dashed) and $M=400$ (solid). Unyielded regions ($\tau \leq 
Bn$) are shown shaded.}
 \label{fig: iso-stress lines}
\end{figure}

An important issue when using the Papanastasiou regularisation is the choice of the exponent $M$. The higher the value of $M$, 
the better the approximation, but also the more difficult it is to solve the equations due to the increased degree of 
nonlinearity. Therefore, for practical reasons $M$ has to be kept within certain limits. Here we provide some comparisons showing 
the effect of $M$ on the quality of the results, while the issue of the computational effort required as a function of $M$ will 
be investigated in the next section. One result of importance in a range of applications is the location of the yield lines, 
which are approximated in the present method by $\tau = Bn$, as previously stated. Figure \ref{fig: yield lines} shows that the 
use of the different values of $100$, $200$ and $400$ for $M$ does not affect the location of the yield lines significantly. The 
difference is especially small between $M=400$ and $M=200$. Figure \ref{fig: iso-stress lines} shows iso-stress lines which 
deviate slightly from the yield stress value. Comparing Figures \ref{fig: yield lines} and \ref{fig: iso-stress lines} shows that 
the lines $\tau=\tau_y+\epsilon$ ($\epsilon > 0$) are less sensitive to the choice of $M$ than the lines $\tau=Bn$. This verifies
the result of Alexandrou and co-workers \cite{burgos_99, burgos_99b} that the location of iso-stress lines $\tau=\tau_y+\epsilon$ 
($\epsilon > 0$) is predicted almost equally well for a range of $M$ values, whereas to accurately predict the location of the 
yield line $\tau=\tau_y$ a high value of $M$ is required. The iso-stress lines $\tau=\tau_y-\epsilon$ ($\epsilon > 0$) nearly 
coincide with the yield stress line except near the side walls and the lower corners of the upper unyielded region. Therefore, 
there is a sharp decrease in stress as one moves into the unyielded zone. On the contrary, this is not observed with the 
$\tau=\tau_y+\epsilon$ lines which are located at a distance from the yield line, meaning that the stress increases more 
gradually as one moves into the yielded zone. It is interesting to note that for $Bn=200$ the difference between the actual 
stress and the yield stress is less than 1\% throughout the fluid region that separates the two solid regions, since the 
$\tau=1.01Bn$ contours do not cross into this region. So, the deformation rate is quite small there, and most of the shear takes 
place close to the lid.

\begin{figure}[t]
\centering
 \includegraphics[scale=0.60]{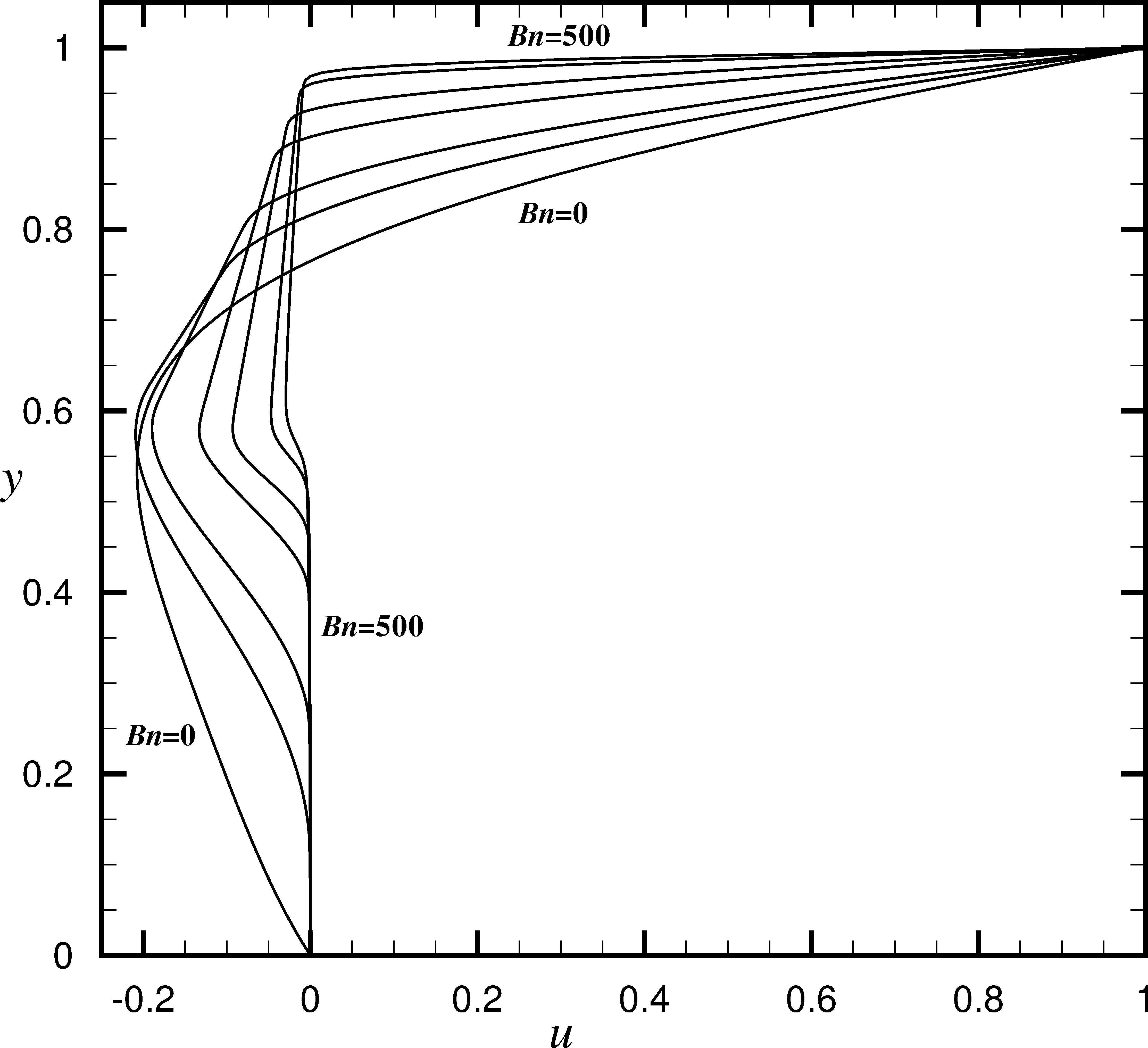}
 \caption{The $x$ component of velocity along the vertical centreline ($x=0.5$), for $Bn$ = $0$, $2$, $5$, $20$, $50$,
$200$ and $500$.}
 \label{fig: centreline U}
\end{figure}

In closing this section we provide some detailed results at the vertical centreline. The $x$ velocity component ($u$) is plotted 
along the vertical centreline in Figure \ref{fig: centreline U}. The limits of the yielded / unyielded zones can be clearly seen 
on that graph. On each curve (except for $Bn=0$) two straight line segments can be identified. One is completely vertical with 
$u=0$ and stretches from $y=0$ up to a certain height; this segment corresponds to the lower unyielded zone which is motionless. 
The other segment corresponds to the upper unyielded zone; there the velocity is non-zero but decreases linearly with height, 
which shows that the upper unyielded zone rotates as a solid body. The centre of rotation can be estimated by extending the 
straight line segment until it intersects the $u=0$ line.

More detailed results are shown in Tables \ref{table: centreline u - Bn=2} ($Bn=2$) and \ref{table: centreline u - Bn=50} 
($Bn=50$), where the velocity values have been obtained at selected points using linear interpolation from the values at adjacent 
CVs. To study grid convergence, results of various grids are also included. The exponent given in the tables is the order of grid 
convergence, which should be $q=2$ for our present second-order accurate scheme. It is calculated by the following formula (see 
\cite{Ferziger_02}):

\begin{equation}
 q \;=\; \frac{\log\left( \frac{u_{128}-u_{64}}{u_{256}-u_{128}} \right)}{\log(2)}
\end{equation}
where the subscripts denote the grid where $u$ has been calculated.

Tables \ref{table: centreline u - Bn=2} and \ref{table: centreline u - Bn=50} show that $q\approx 1.5$ for $Bn=2$, and $q\approx 
1$ for $Bn=50$. So, second-order convergence is not exhibited, which could be a sign that an even finer grid would be useful. 
However, the discretisation error is already relatively small as can be seen from the difference between the solutions at various 
grids. Besides, the use of a finer grid would significantly increase the computational cost as will be described in the next 
section. For $Bn=50$ the value of $q$ is especially low at the upper parts of the centreline, which is an indication that at 
higher $Bn$ numbers (when most of the flow occurs at the upper part of the domain) it would be of benefit to use non-uniform 
grids with increased resolution near the top, or adaptive grids.

The last three columns of the tables are defined as follows:

\begin{equation}
 \delta_{256}^{128} = 100 \!\cdot\! \left| \frac{u_{256}^{400}-u_{128}^{400}}{u_{256}^{400}} \right| \%, \quad
 \delta_{400}^{200} = 100 \!\cdot\! \left| \frac{u_{256}^{400}-u_{256}^{200}}{u_{256}^{400}} \right| \%, \quad
 \delta_{400}^{100} = 100 \!\cdot\! \left| \frac{u_{256}^{400}-u_{256}^{100}}{u_{256}^{400}} \right| \%
\end{equation}
where $u_n^m$ has been calculated on grid $n\times n$ with $M=m$. These values must be interpreted with caution, as they may 
appear large if $u_{256}^{400}$ is close to zero. $\delta_{256}^{128}$ is a measure of the discretisation error, and it can be 
noticed that for $Bn = 2$ it is quite small, of the order of $0.1$--$1$\%, whereas for $Bn = 50$ it is markedly larger, of the 
order of $5$\%. For $Bn = 2$, $\delta_{400}^{200}$ is of the order of $0.1$\%, while $\delta_{400}^{100}$ is about $3$ times 
higher at each point. $\delta_{400}^{200}$ shows greater variability for $Bn = 50$, and is a little higher than for $Bn = 2$. 
$\delta_{400}^{100}$, for $Bn = 50$, has about twice the value of $\delta_{400}^{200}$ in the upper part of the centreline, and 
three times the value of $\delta_{400}^{200}$ in the lower part. In general, it appears that, with the present choices of grid 
density and $M$, grid coarseness is a larger source of error than the smallness of $M$.

\begin{table}[p]
\caption{Values of the $x$-component of velocity along the vertical centreline for $Bn = 2$.}
\label{table: centreline u - Bn=2}
\begin{center}
\begin{scriptsize}   % to make the font smaller
% Alternating grey/white rows starting at row 3:
%\rowcolors{4}{black!20}{white}
\renewcommand\arraystretch{1.0}   % Make row height smaller (default=1)
\begin{tabular}{ c | r r | r | r r || r || r r r }
 \hline
  grid & $64 \times 64$ & $128 \times 128$ & $256 \times 256$ &  $256 \times 256$ & $256 \times 256$ &
  \multirow{2}{*}{\textbf{\textit{q}}} &
  \multirow{2}{*}{$\delta_{256}^{128}$} &
  \multirow{2}{*}{$\delta_{400}^{200}$} &  
  \multirow{2}{*}{$\delta_{400}^{100}$}
\\ \cline{1-1}
   $y$ & $M=400$        & $M=400$          & $M=400$          & $M=200$           & $M=100$          & & & &
  \\
 \hline
 \hline
1.00  &  1.00000 &  1.00000 &  1.00000 &  1.00000 &  1.00000 &    - &  0.00\% &  0.00\%  &   0.00\% \\
0.975 &  0.81421 &  0.81544 &  0.81581 &  0.81581 &  0.81581 & 1.74 &  0.05\% &  0.00\%  &   0.00\% \\
0.950 &  0.63974 &  0.64222 &  0.64279 &  0.64280 &  0.64282 & 2.13 &  0.09\% &  0.00\%  &   0.01\% \\
0.900 &  0.33677 &  0.34016 &  0.34121 &  0.34126 &  0.34136 & 1.69 &  0.31\% &  0.02\%  &   0.04\% \\
0.850 &  0.10671 &  0.11033 &  0.11160 &  0.11172 &  0.11193 & 1.51 &  1.14\% &  0.11\%  &   0.29\% \\
0.800 & -0.04182 & -0.03945 & -0.03834 & -0.03811 & -0.03774 & 1.09 &  2.90\% &  0.59\%  &   1.56\% \\
0.750 & -0.10310 & -0.10602 & -0.10703 & -0.10688 & -0.10653 & 1.53 &  0.95\% &  0.14\%  &   0.47\% \\
0.700 & -0.13840 & -0.14159 & -0.14267 & -0.14260 & -0.14242 & 1.56 &  0.76\% &  0.05\%  &   0.18\% \\
0.650 & -0.17334 & -0.17689 & -0.17810 & -0.17793 & -0.17757 & 1.56 &  0.68\% &  0.09\%  &   0.29\% \\
0.600 & -0.20322 & -0.20575 & -0.20606 & -0.20574 & -0.20516 & 3.05 &  0.15\% &  0.16\%  &   0.44\% \\
0.550 & -0.20660 & -0.20681 & -0.20679 & -0.20654 & -0.20610 &    - &  0.01\% &  0.12\%  &   0.34\% \\
0.500 & -0.18918 & -0.18905 & -0.18898 & -0.18878 & -0.18841 & 0.82 &  0.04\% &  0.11\%  &   0.30\% \\
0.450 & -0.16059 & -0.16039 & -0.16033 & -0.16018 & -0.15990 & 1.89 &  0.03\% &  0.09\%  &   0.27\% \\
0.400 & -0.12699 & -0.12678 & -0.12674 & -0.12665 & -0.12647 & 2.45 &  0.03\% &  0.07\%  &   0.21\% \\
0.350 & -0.09272 & -0.09260 & -0.09261 & -0.09259 & -0.09255 &    - &  0.02\% &  0.03\%  &   0.08\% \\
0.300 & -0.06101 & -0.06112 & -0.06120 & -0.06126 & -0.06138 & 0.23 &  0.14\% &  0.10\%  &   0.28\% \\
0.250 & -0.03421 & -0.03464 & -0.03485 & -0.03502 & -0.03534 & 1.07 &  0.59\% &  0.48\%  &   1.40\% \\
0.200 & -0.01443 & -0.01493 & -0.01523 & -0.01554 & -0.01613 & 0.78 &  1.92\% &  2.05\%  &   5.94\% \\
0.150 & -0.00365 & -0.00344 & -0.00356 & -0.00408 & -0.00508 &    - &  3.21\% & 14.68\%  &  42.93\% \\
0.100 & -0.00078 & -0.00063 & -0.00058 & -0.00111 & -0.00208 & 1.76 &  7.88\% & 90.21\%  & 257.00\% \\
0.050 & -0.00030 & -0.00028 & -0.00027 & -0.00053 & -0.00101 & 1.64 &  2.42\% & 94.50\%  & 272.53\% \\
0.000 &  0.00000 &  0.00000 &  0.00000 &  0.00000 &  0.00000 &    - &  0.00\% &  0.00\%  &   0.00\% \\
 \hline
\end{tabular}
\end{scriptsize}
\end{center}
\end{table}

\begin{table}[p]
\caption{Values of the $x$-component of velocity along the vertical centreline for $Bn = 50$.}
\label{table: centreline u - Bn=50}
\begin{center}
\begin{scriptsize}   % to make the font smaller
% Alternating grey/white rows starting at row 3:
%\rowcolors{4}{black!20}{white}
\renewcommand\arraystretch{1.0}   % Make row height smaller (default=1)
\begin{tabular}{ c | r r | r | r r || r || r r r }
 \hline
  grid & $64 \times 64$ & $128 \times 128$ & $256 \times 256$ &  $256 \times 256$ & $256 \times 256$ &
  \multirow{2}{*}{\textbf{\textit{q}}} &
  \multirow{2}{*}{$\delta_{256}^{128}$} &
  \multirow{2}{*}{$\delta_{400}^{200}$} &  
  \multirow{2}{*}{$\delta_{400}^{100}$}
\\ \cline{1-1}
   $y$ & $M=400$        & $M=400$          & $M=400$          & $M=200$           & $M=100$          & & & &
  \\
 \hline
 \hline
1.0000 &  1.00000 &  1.00000 &  1.00000 &  1.00000 &  1.00000 &     - &  0.00\% &   0.00\%  &   0.00\% \\
0.9875 &  0.71205 &  0.72144 &  0.72838 &  0.72886 &  0.72931 &  0.43 &  0.95\% &   0.07\%  &   0.13\% \\
0.9750 &  0.45957 &  0.47518 &  0.49123 &  0.49225 &  0.49324 & -0.04 &  3.27\% &   0.21\%  &   0.41\% \\
0.9500 &  0.10907 &  0.11660 &  0.13617 &  0.13835 &  0.14067 & -1.38 & 14.38\% &   1.60\%  &   3.30\% \\
0.9250 & -0.01230 & -0.01713 & -0.01946 & -0.01812 & -0.01641 &  1.05 & 11.97\% &   6.90\%  &  15.70\% \\
0.9000 & -0.02462 & -0.02952 & -0.03123 & -0.03106 & -0.03073 &  1.52 &  5.49\% &   0.56\%  &   1.60\% \\
0.8500 & -0.03500 & -0.03941 & -0.04142 & -0.04143 & -0.04142 &  1.14 &  4.84\% &   0.03\%  &   0.01\% \\
0.8000 & -0.04469 & -0.04914 & -0.05148 & -0.05156 & -0.05165 &  0.93 &  4.55\% &   0.16\%  &   0.33\% \\
0.7500 & -0.05431 & -0.05880 & -0.06148 & -0.06157 & -0.06162 &  0.75 &  4.36\% &   0.14\%  &   0.23\% \\
0.7000 & -0.06386 & -0.06840 & -0.07142 & -0.07145 & -0.07135 &  0.59 &  4.23\% &   0.04\%  &   0.09\% \\
0.6500 & -0.07328 & -0.07788 & -0.08125 & -0.08113 & -0.08069 &  0.45 &  4.15\% &   0.15\%  &   0.69\% \\
0.6000 & -0.08172 & -0.08678 & -0.09080 & -0.09033 & -0.08917 &  0.33 &  4.43\% &   0.52\%  &   1.80\% \\
0.5750 & -0.08318 & -0.08778 & -0.09275 & -0.09180 & -0.08990 & -0.11 &  5.36\% &   1.03\%  &   3.08\% \\
0.5500 & -0.07734 & -0.07753 & -0.08027 & -0.07921 & -0.07706 & -3.91 &  3.42\% &   1.32\%  &   3.99\% \\
0.5250 & -0.06148 & -0.05345 & -0.05286 & -0.05252 & -0.05126 &  3.76 &  1.12\% &   0.63\%  &   3.01\% \\
0.5000 & -0.03959 & -0.02641 & -0.02406 & -0.02464 & -0.02484 &  2.49 &  9.75\% &   2.41\%  &   3.23\% \\
0.4500 & -0.00874 & -0.00255 & -0.00137 & -0.00233 & -0.00408 &  2.38 & 86.69\% &  70.45\%  & 198.24\% \\
0.4000 & -0.00153 & -0.00076 & -0.00069 & -0.00135 & -0.00263 &  3.46 & 10.19\% &  96.74\%  & 283.34\% \\
0.3000 & -0.00048 & -0.00040 & -0.00038 & -0.00075 & -0.00146 &  1.95 &  5.70\% &  97.59\%  & 287.10\% \\
0.2000 & -0.00026 & -0.00022 & -0.00021 & -0.00042 & -0.00083 &  1.99 &  4.82\% &  97.74\%  & 287.60\% \\
0.1000 & -0.00013 & -0.00011 & -0.00010 & -0.00020 & -0.00040 &  2.08 &  4.40\% &  97.74\%  & 287.34\% \\
0.0000 &  0.00000 &  0.00000 &  0.00000 &  0.00000 &  0.00000 &     - &  0.00\% &   0.00\%  &   0.00\% \\
 \hline
\end{tabular}
\end{scriptsize}
\end{center}
\end{table}

\section{Convergence of the algebraic solver}
\label{sec: algebraic convergence}

When using the SIMPLE algorithm to solve Bingham flows, either as a single-grid solver or in a multigrid context, one realises 
that there is a severe deterioration of the convergence rates as the Bingham number increases. In Figure \ref{fig: SG convergence 
per Bn} we plot against the number of iterations the $L^{\infty}$-norm of the residual vector of the $x-$momentum equations,

\begin{equation} \label{eq: residual norm}
 \lVert r \rVert_{\infty} \;=\; \max_{P=1,\ldots,N} \left\{|r_P|\right\} \;,
\end{equation}
where $r_P$ is the residual, expressed per unit volume, of the $x-$momentum equation of control volume $P$ and $N$ is the total 
number of control volumes in the grid. The residual norm is scaled by 1000 because in the actual numerical experiments we solved 
the dimensional version of the equations, with $L=1$ m, $U=1$ m/s, and $\mu=1000$ Pa$\cdot$s. Therefore, the residuals of the 
dimensional equations are 1000 times larger than those of the dimensionless equations.

\begin{figure}[htb]
\centering
 \includegraphics[scale=0.6]{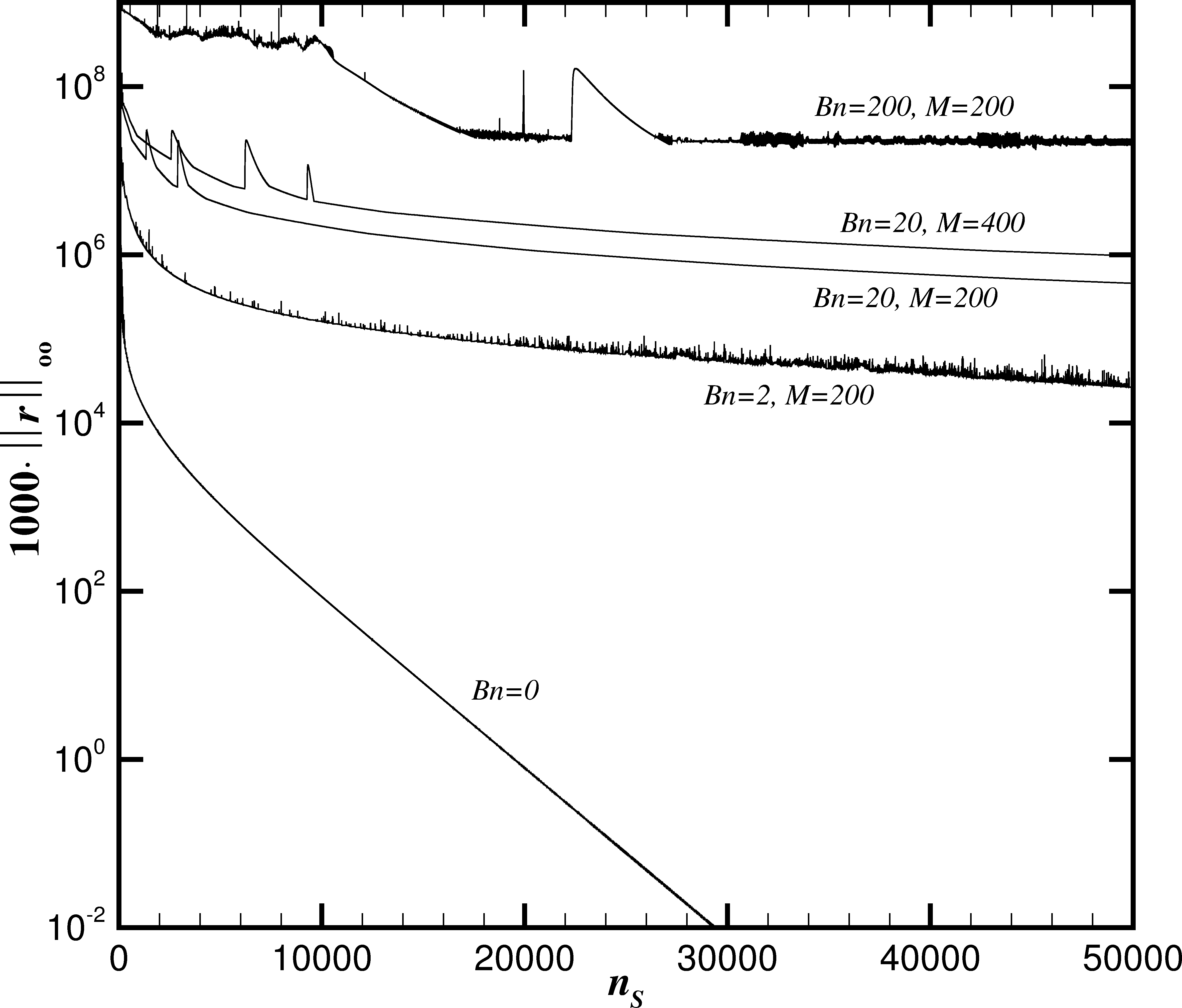}
 \caption{The $L^{\infty}$ norm of the $x$-momentum residual plotted against the number $n_S$ of SIMPLE iterations, for
single grid solution; $a_u=0.7$, $a_p=0.3$, $256\times 256$ grid.}
 \label{fig: SG convergence per Bn}
\end{figure}

\begin{figure}[htb]
\centering
 \includegraphics[scale=0.6]{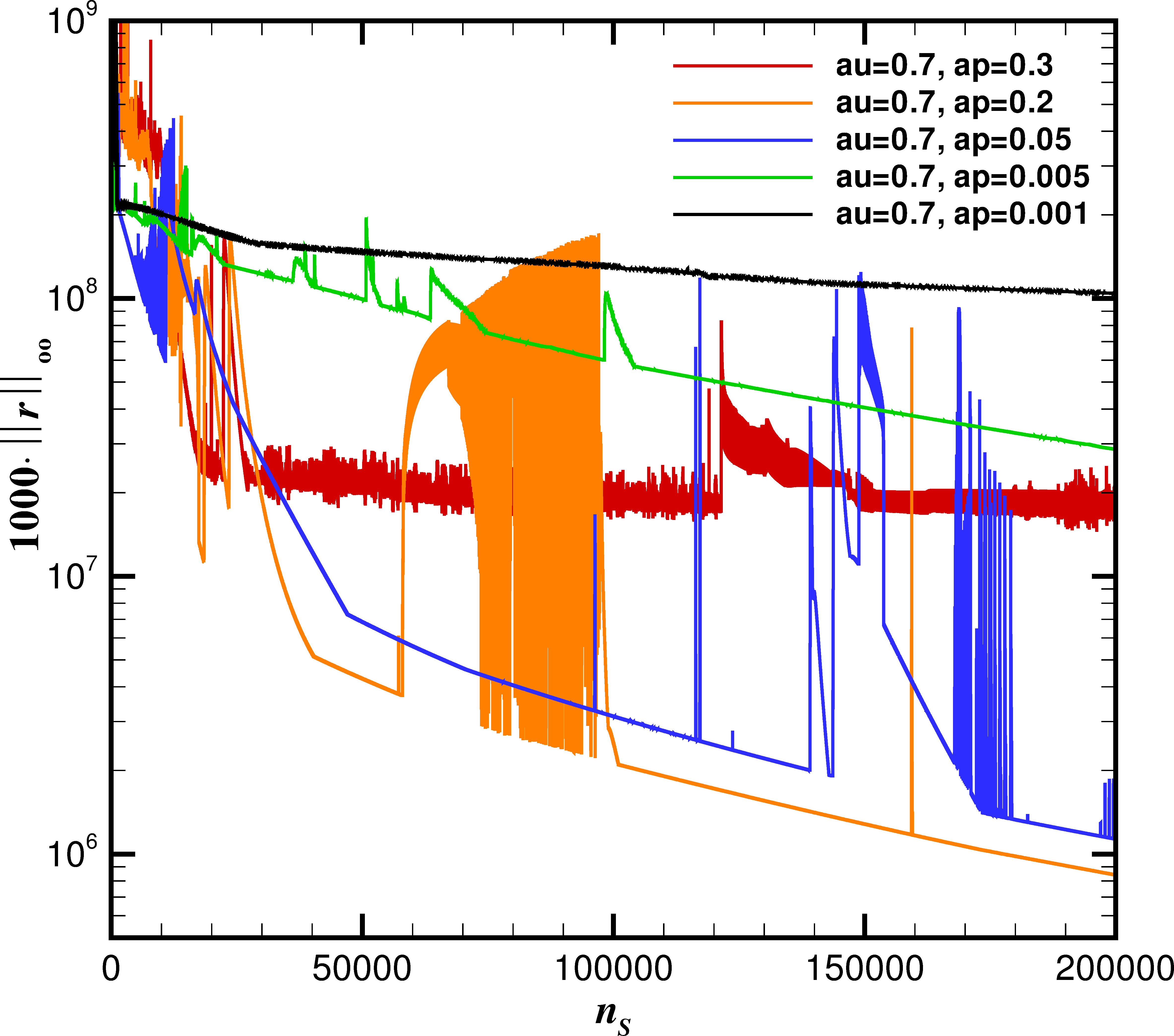}
 \caption{The $L^{\infty}$ norm of the $x$-momentum residual plotted against the number $n_S$ of SIMPLE iterations, for
single grid solution; $Bn=200$, $M=200$, $256\times 256$ grid.}
 \label{fig: SG convergence per ap}
\end{figure}

Figure \ref{fig: SG convergence per Bn} shows that the performance of SIMPLE, as a single-grid solver, deteriorates as either 
$Bn$ or $M$ increases. The choices $a_u=0.7$ and $a_p=0.3$, used to obtain these results, are reasonable for Newtonian flows 
\cite{Ferziger_02}. In Figure \ref{fig: SG convergence per ap}, the effect of varying $a_p$ is illustrated. As $a_p$ is 
increased, convergence becomes faster, especially at the initial stages of iteration; however, it also becomes more oscillatory 
with large spikes. Beyond a certain value of $a_p$ ($a_p=0.3$ in Figure \ref{fig: SG convergence per ap}) convergence stalls. On 
the other hand, using a very small $a_p$ results in smooth but slow convergence. Due to the very large number of combinations of 
$Bn$, $M$, $a_u$, $a_p$ and grid density, it is impossible to investigate fully the effect of all these parameters. We have 
observed (results not shown) that the performance of SIMPLE is similar also for other values of $a_u$ in the range $0.8$ - $0.2$, 
but beyond these values it is difficult to obtain convergence.

\begin{figure}[p]
\centering
 \includegraphics[scale=0.55]{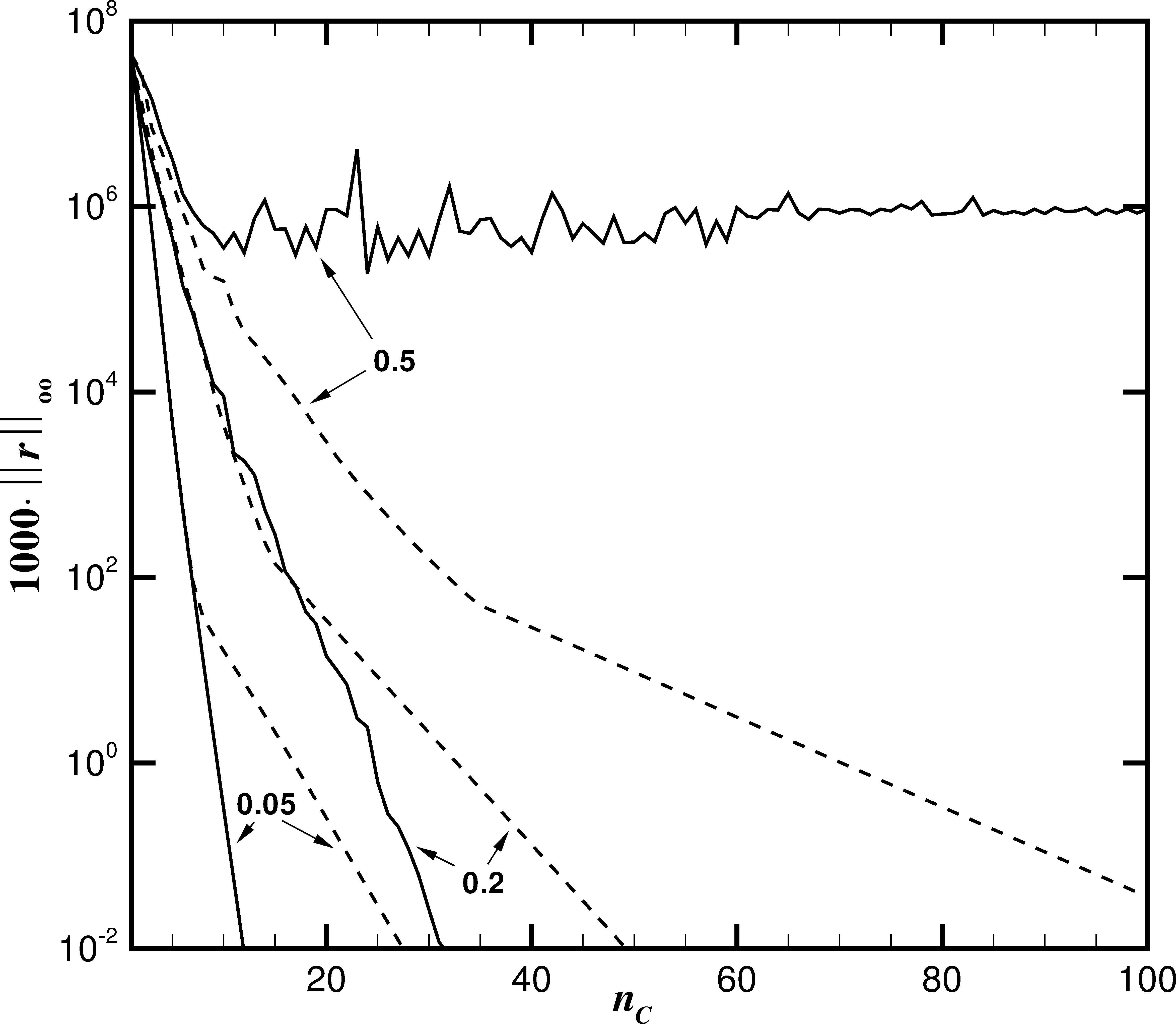}
 \caption{The $L^{\infty}$ norm of the $x$-momentum residual plotted against the number of multigrid cycles $n_C$, for
$Bn=0.05$, $0.2$, and $0.5$ ($M=400$, grid $256\times 256$). Solid lines depict convergence of the standard multigrid
method, while dashed lines depict convergence of the modified multigrid method where viscosity is not updated at the
coarse grids, but interpolated from the immediately finer grid.}
 \label{fig: classic MG convergence - grid 256}
\end{figure}

The performance of SIMPLE as a smoother in a multigrid context is tested next. In applying multigrid, the coarsest grid used in 
the cycles was the $8 \times 8$ grid. According to our numerical experiments, the convergence deterioration of the SIMPLE 
algorithm with increasing $Bn$ number also reflects on the multigrid performance. In fact, as Figure \ref{fig: classic MG 
convergence - grid 256} shows (solid lines), the multigrid method does not converge beyond a relatively small value of $Bn$, 
around 0.5. As a remedy, we applied the suggestion of Ferziger and Peric \cite{Ferziger_02}, that if a fluid property, such as 
the viscosity when RANS turbulence models are used, varies by orders of magnitude within the computational domain, then it may be 
useful to update that property only on the finest grid and keep it constant within a multigrid cycle. So, we tried an 
implementation where the viscosity at coarse grids is not calculated afresh from (\ref{eq: papanastasiou_eta nd}) according to 
the restricted velocity field, but it is directly restricted (interpolated) from the immediately finer grid. The convergence of 
this modified multigrid method is also shown in Figure \ref{fig: classic MG convergence - grid 256} (dashed lines). The method 
converges more slowly than the standard multigrid method, but it is more robust and can converge over a wider range of Bingham 
numbers. Both methods converge equally fast up to a point, beyond which the convergence of the modified method suddenly slows 
down. It appears that both methods are equally capable of reducing certain components of the residual, which dominate the 
residual at the initial stages of iteration. However, the modified method is less capable of reducing certain other components, 
which dominate the error beyond a certain point.

\begin{figure}[p]
\centering
 \includegraphics[scale=0.55]{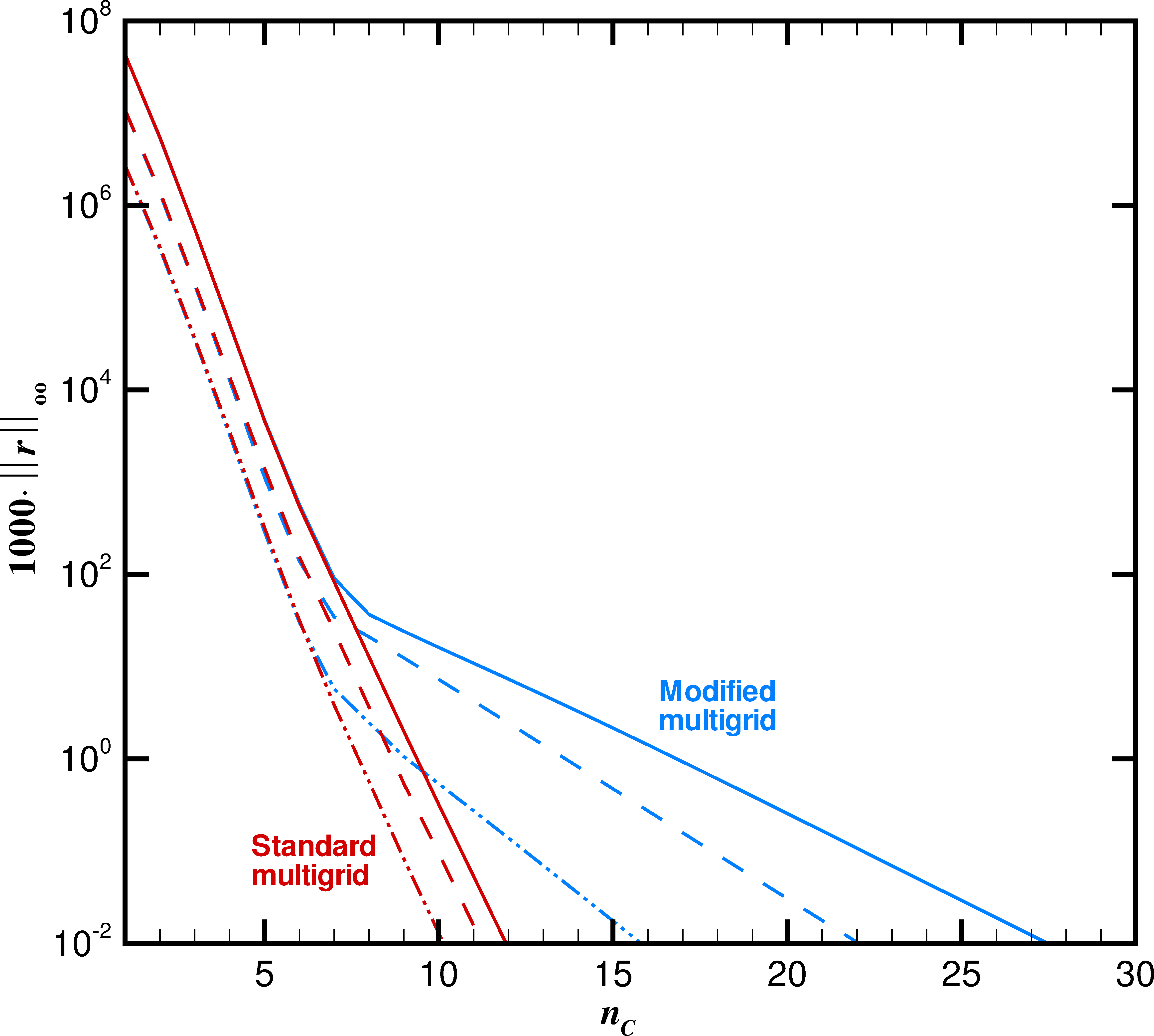}
 \caption{The $L^{\infty}$ norm of the $x$-momentum residual plotted against the number of multigrid cycles, $n_C$, for
$Bn=0.05$, $M=400$, on various grids: $64\times 64$ (chained lines), $128\times 128$ (dashed lines), and $256\times
256$ (solid lines). In the modified multigrid method, viscosity is not updated at the coarse grids, but simply
interpolated from the immediately finer grid.}
 \label{fig: classic MG convergence - Bn=0.05}
\end{figure}

The modified method is not strictly ``multigrid'' since part of the solution, namely the updating of the viscosity, only takes 
place on the finest grid. Therefore the method exhibits also some single-grid convergence characteristics. In particular, as 
Figure \ref{fig: classic MG convergence - Bn=0.05} shows for $Bn=0.05$, the convergence of the modified method slows down as the 
grid is refined (this is true only for those slowly converging components of the residual). On the contrary, the standard method 
converges equally fast on all grids, which is normal multigrid behaviour.

For higher Bingham numbers it is necessary to use the modified multigrid method, and even in that case convergence difficulties 
are encountered. However, the gains compared to the single-grid method are still quite impressive. Figure \ref{fig: MG vs SG 
convergence} shows the convergence rates for various Bingham numbers, using both single-grid (SG) and multigrid (MG) procedures. 
In each case the solution on the $128 \times 128$ grid was used as the initial guess and the coarsest grid used by the multigrid 
cycles was the $8 \times 8$ grid. The computational effort ($x$--axis) is measured in terms of equivalent fine-grid SIMPLE 
iterations. For the single-grid cases, this is just the number of SIMPLE iterations performed. For the multigrid cases, the 
number of cycles is multiplied by the number of fine-grid SIMPLE iterations that cost computationally the same as a single cycle. 
For example, $n_C$ W($\nu_1,\nu_2$)--$\nu_3$ cycles cost approximately the same as $n_S = n_C\cdot[2(\nu_1+\nu_2)+\nu_3]$ SIMPLE 
iterations on the finest grid, as mentioned in section \ref{sec: solution}. It should be noted that the cost of restriction and 
prolongation is omitted in this calculation, since the cost of these operations is very small compared to the cost of the SIMPLE 
iterations, especially if one considers that the numbers of pre- and post- smoothing iterations are large, and fine-grid 
iterations are also carried out between cycles. Therefore, MG and SG convergence rates are directly comparable in Fig. \ref{fig: 
MG vs SG convergence}.

\begin{figure}[htb]
\centering
\noindent\makebox[\textwidth]{
 \subfigure []{\label{sfig: MG convergence, far}
  \includegraphics[scale=0.45]{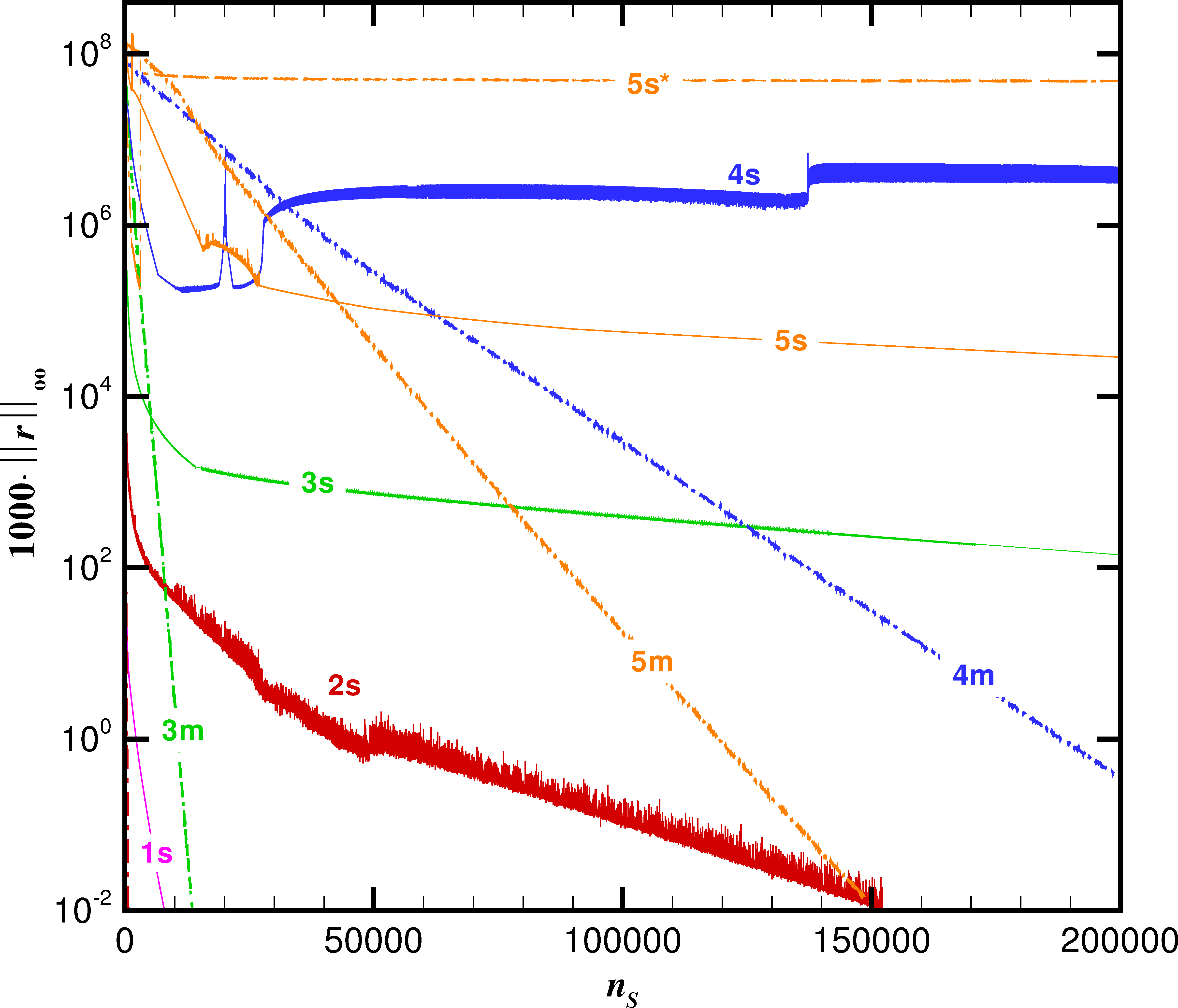}}
 \subfigure []{\label{sfig: MG convergence, close}
  \includegraphics[scale=0.45]{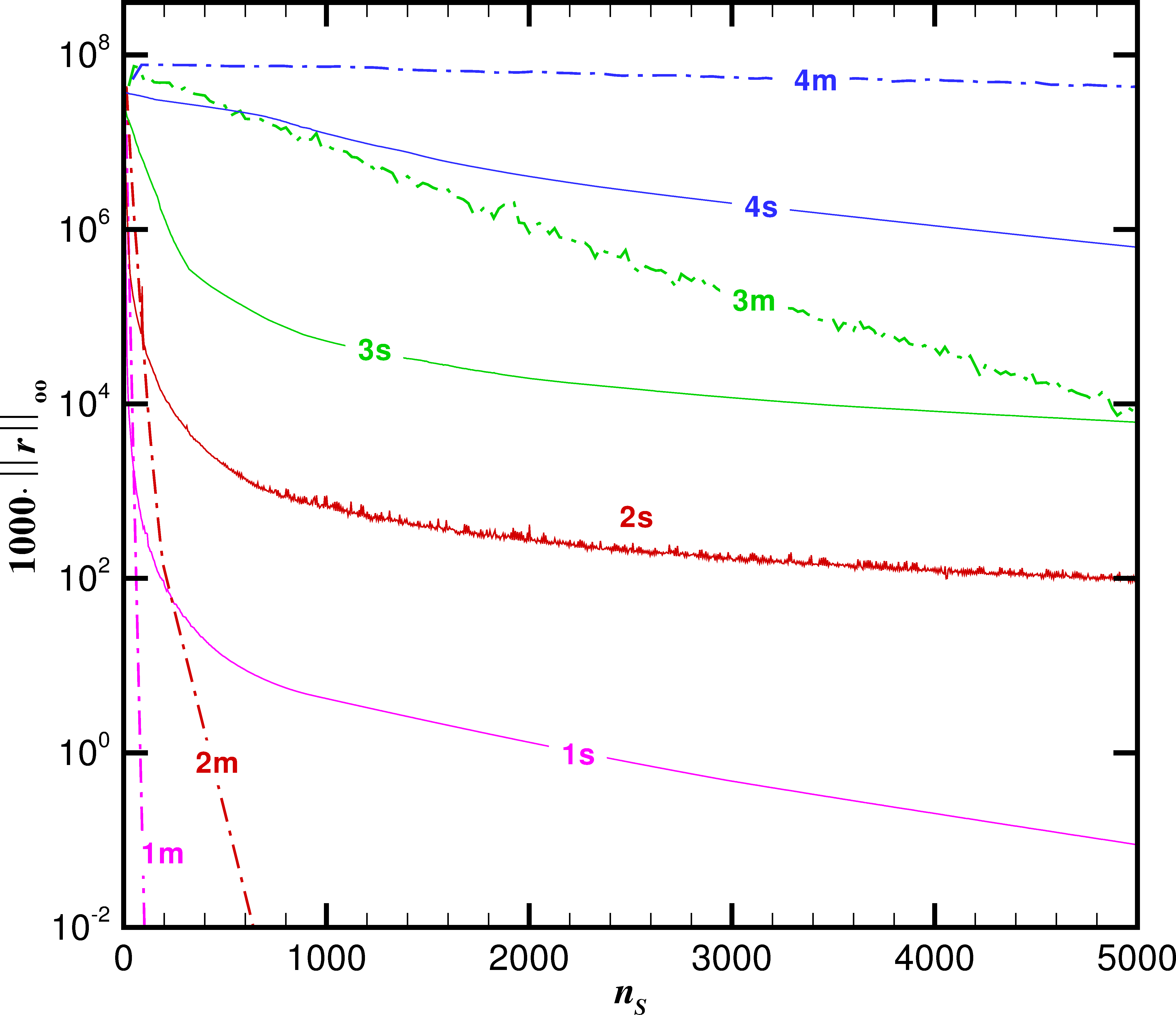}}
}
 \includegraphics[scale=0.55]{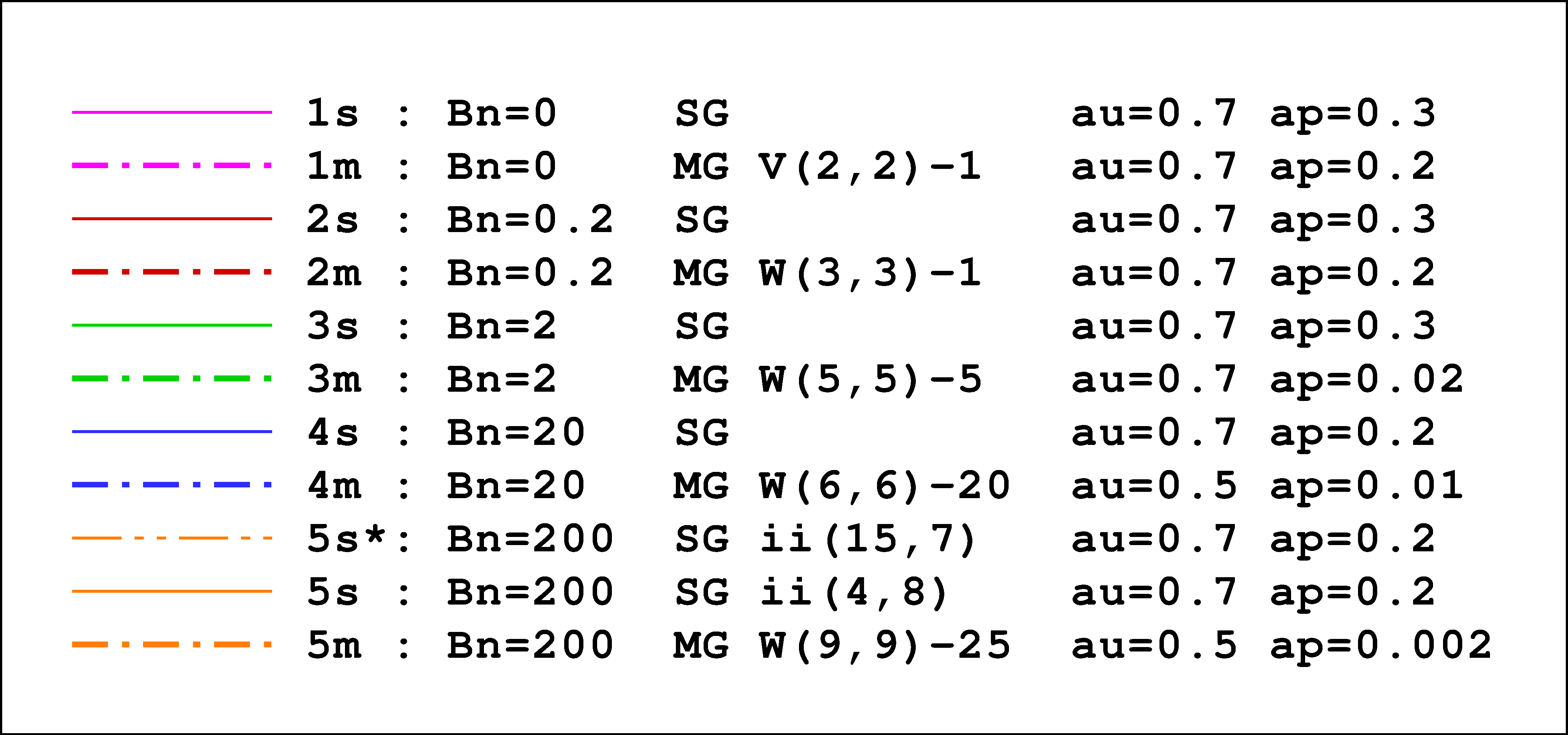}
 \caption{\subref{sfig: MG convergence, far} The $L^{\infty}$ norm of the $x$-momentum residual on the $256\times 256$
grid plotted against the number of equivalent fine-grid SIMPLE iterations, $n_S$. In all cases $M=400$, except $M=200$
for $Bn=200$. SG and MG denote respectively single-grid and multigrid solution procedures, the initial guess being the
solution of the immediately finer ($128\times 128$) grid. \subref{sfig: MG convergence, close} Zoom--in of \subref{sfig:
MG convergence, far} so that the curves \texttt{1m} and \texttt{2m} (which are too close to the vertical axis to be
visible in \subref{sfig: MG convergence, far}) are visible. Also, to improve the readability, the results for $Bn=200$
have been omitted from graph \subref{sfig: MG convergence, close}.}
 \label{fig: MG vs SG convergence}
\end{figure}

One may also observe that as the $Bn$ number increases, the choice of multigrid parameters becomes rather unusual compared to the 
usual multigrid practice: the number of pre- and post-smoothing sweeps is quite large, and a large number of SIMPLE iterations 
are required between cycles. These choices of parameters have been found necessary to obtain convergence. It may also be seen 
that very small values of $a_p$ were used in the multigrid cases. This was necessary, otherwise the procedure did not converge. 
This may be associated with the fact that, as $a_p$ increases, SIMPLE converges faster but in a very oscillatory manner, which 
may cause problems to the multigrid procedure. Since the single-grid procedure converges faster when $a_p$ is as large as 
possible, the SG results have been obtained with large values of $a_p$, compared with the MG cases. It can be observed that 
multigrid is extremely efficient at low $Bn$ numbers, but has difficulty when $Bn$ is large. Yet, in every case the multigrid 
convergence rates are much faster than the single-grid rates, except for the first few iterations. Therefore multigrid is 
preferable in any case. Moreover, we note that multigrid (with the particular choices of parameters) converges faster for 
$Bn=200$ with $M=200$ than for $Bn=20$ with $M = 400$. Indeed we observed in every case that convergence deteriorates 
significantly not only with increasing $Bn$ but also with increasing $M$ (for every Bingham number we solved the problem using 
$M=100$, $200$, and $400$, except for $Bn=500$ and $1000$ where only $M=100$ was used due to convergence difficulties).

It is worth mentioning that the number of inner iterations seems to play an important role. In Figure \ref{fig: MG vs SG 
convergence} it can be seen that for $Bn=200$, using $4$ GMRES iterations and $8$ CG iterations for the velocity and pressure 
correction systems (labelled ``ii(4,8)'' in the figure) significantly improves the convergence rate of the single-grid procedure 
compared to the case that $15$ and $7$ iterations are used instead (labelled ``ii(15,7)''). It seems that due to the nonlinearity 
of the algebraic system it is better not to perform many iterations on the linearised velocity systems within each SIMPLE outer 
iteration. This was not investigated in detail, due to the fact that there are already a large number of parameters involved in 
the solution procedure.

\section{Conclusions}
\label{sec:conclusions}

We have solved the creeping lid-driven cavity flow of a Bingham plastic using the Papanastasiou regularization and the finite 
volume method combined with a multigrid algorithm. Results have been obtained for Bingham numbers in the range 0--1000. These 
compare favorably with the results of other methods, such as the finite-element and the finite-difference method, and show that 
the proposed method provides a useful tool in solving viscoplastic flows for a wide range of Bingham numbers. It should be noted 
that the convergence of the method becomes slow at high values of the Bingham number and the regularization parameter. With the 
use of a modified multigrid method convergence is accelerated considerably compared to the single-grid SIMPLE method.

\section*{Acknowledgements}
This work was co-funded by the European Regional Development fund and the Republic of Cyprus through the Research Promotion 
Foundation (research project $\mathrm{AEI\Phi OPIA/\Phi Y\Sigma H}$/0609(BIE)/15).



%% References


% \end{linenumbers}

% \clearpage
\section*{REFERENCES}
\bibliographystyle{model1-num-names}
\bibliography{bingham.bib}

\begin{thebibliography}{40}
\expandafter\ifx\csname natexlab\endcsname\relax\def\natexlab#1{#1}\fi
\providecommand{\bibinfo}[2]{#2}
\ifx\xfnm\relax \def\xfnm[#1]{\unskip,\space#1}\fi
%Type = Article
\bibitem[{Barnes(1999)}]{Barnes_99}
\bibinfo{author}{H.~A. Barnes},
\newblock \bibinfo{title}{The yield stress -- a review or
  `$\pi\alpha\nu\tau\alpha$ $\rho\epsilon\iota$' -- everything flows?},
\newblock \bibinfo{journal}{Journal of Non-Newtonian Fluid Mechanics}
  \bibinfo{volume}{81} (\bibinfo{year}{1999}) \bibinfo{pages}{133 -- 178}.
%Type = Book
\bibitem[{Bingham(1922)}]{Bingham_22}
\bibinfo{author}{E.~C. Bingham}, \bibinfo{title}{Fluidity and plasticity},
  \bibinfo{publisher}{McGraw-Hill, New York}, \bibinfo{year}{1922}.
%Type = Article
\bibitem[{Papanastasiou(1987)}]{papanastasiou_87}
\bibinfo{author}{T.~C. Papanastasiou},
\newblock \bibinfo{title}{Flows of materials with yield},
\newblock \bibinfo{journal}{Journal of Rheology} \bibinfo{volume}{31}
  (\bibinfo{year}{1987}) \bibinfo{pages}{385--404}.
%Type = Article
\bibitem[{Frigaard and Nouar(2005)}]{Frigaard_05}
\bibinfo{author}{I.~Frigaard}, \bibinfo{author}{C.~Nouar},
\newblock \bibinfo{title}{On the usage of viscosity regularisation methods for
  visco-plastic fluid flow computation},
\newblock \bibinfo{journal}{Journal of Non-Newtonian Fluid Mechanics}
  \bibinfo{volume}{127} (\bibinfo{year}{2005}) \bibinfo{pages}{1 -- 26}.
%Type = Article
\bibitem[{O'Donovan and Tanner(1984)}]{Donovan_84}
\bibinfo{author}{E.~O'Donovan}, \bibinfo{author}{R.~Tanner},
\newblock \bibinfo{title}{Numerical study of the {B}ingham squeeze film
  problem},
\newblock \bibinfo{journal}{Journal of Non-Newtonian Fluid Mechanics}
  \bibinfo{volume}{15} (\bibinfo{year}{1984}) \bibinfo{pages}{75 -- 83}.
%Type = Book
\bibitem[{Fortin and Glowinski(1983)}]{Fortin_83}
\bibinfo{author}{M.~Fortin}, \bibinfo{author}{R.~Glowinski},
  \bibinfo{title}{{A}ugmented {L}agrangian {M}ethods: applications to the
  numerical solution of boundary-value problems},
  \bibinfo{publisher}{North-Holland, Amsterdam}, \bibinfo{year}{1983}.
%Type = Book
\bibitem[{Glowinski(1984)}]{Glowinski_84}
\bibinfo{author}{R.~Glowinski}, \bibinfo{title}{Numerical methods for nonlinear
  variational problems}, \bibinfo{publisher}{Springer, New York},
  \bibinfo{year}{1984}.
%Type = Article
\bibitem[{Dean et~al.(2007)Dean, Glowinski, and Guidoboni}]{Dean_07}
\bibinfo{author}{E.~J. Dean}, \bibinfo{author}{R.~Glowinski},
  \bibinfo{author}{G.~Guidoboni},
\newblock \bibinfo{title}{On the numerical simulation of {B}ingham
  visco-plastic flow: Old and new results},
\newblock \bibinfo{journal}{Journal of Non-Newtonian Fluid Mechanics}
  \bibinfo{volume}{142} (\bibinfo{year}{2007}) \bibinfo{pages}{36 -- 62}.
%Type = Article
\bibitem[{Neofytou(2005)}]{neofytou_05}
\bibinfo{author}{P.~Neofytou},
\newblock \bibinfo{title}{A 3rd order upwind finite volume method for
  generalised {N}ewtonian fluid flows},
\newblock \bibinfo{journal}{Advances in Engineering Software}
  \bibinfo{volume}{36} (\bibinfo{year}{2005}) \bibinfo{pages}{664--680}.
%Type = Article
\bibitem[{Patankar and Spalding(1972)}]{Patankar_72}
\bibinfo{author}{S.~V. Patankar}, \bibinfo{author}{D.~B. Spalding},
\newblock \bibinfo{title}{A calculation procedure for heat, mass and momentum
  transfer in three-dimensional parabolic flows},
\newblock \bibinfo{journal}{International Journal of Heat and Mass Transfer}
  \bibinfo{volume}{15} (\bibinfo{year}{1972}) \bibinfo{pages}{1787--1806}.
%Type = Article
\bibitem[{Turan et~al.(2010)Turan, Chakraborty, and Poole}]{Turan_10}
\bibinfo{author}{O.~Turan}, \bibinfo{author}{N.~Chakraborty},
  \bibinfo{author}{R.~J. Poole},
\newblock \bibinfo{title}{Laminar natural convection of {B}ingham fluids in a
  square enclosure with differentially heated side walls},
\newblock \bibinfo{journal}{Journal of Non-Newtonian Fluid Mechanics}
  \bibinfo{volume}{165} (\bibinfo{year}{2010}) \bibinfo{pages}{901 -- 913}.
%Type = Article
\bibitem[{Turan et~al.(2012)Turan, Chakraborty, and Poole}]{Turan_12}
\bibinfo{author}{O.~Turan}, \bibinfo{author}{N.~Chakraborty},
  \bibinfo{author}{R.~J. Poole},
\newblock \bibinfo{title}{Laminar {R}ayleigh-{B}enard convection of yield
  stress fluids in a square enclosure},
\newblock \bibinfo{journal}{Journal of Non-Newtonian Fluid Mechanics}
  \bibinfo{volume}{171-172} (\bibinfo{year}{2012}) \bibinfo{pages}{83 -- 96}.
%Type = Article
\bibitem[{Neofytou and Drikakis(2003{\natexlab{a}})}]{Neofytou_03}
\bibinfo{author}{P.~Neofytou}, \bibinfo{author}{D.~Drikakis},
\newblock \bibinfo{title}{Effects of blood models on flows through a stenosis},
\newblock \bibinfo{journal}{International Journal for Numerical Methods in
  Fluids} \bibinfo{volume}{43} (\bibinfo{year}{2003}{\natexlab{a}})
  \bibinfo{pages}{597--635}.
%Type = Article
\bibitem[{Neofytou and Drikakis(2003{\natexlab{b}})}]{Neofytou_03b}
\bibinfo{author}{P.~Neofytou}, \bibinfo{author}{D.~Drikakis},
\newblock \bibinfo{title}{Non-{N}ewtonian flow instability in a channel with a
  sudden expansion},
\newblock \bibinfo{journal}{Journal of Non-Newtonian Fluid Mechanics}
  \bibinfo{volume}{111} (\bibinfo{year}{2003}{\natexlab{b}})
  \bibinfo{pages}{127 -- 150}.
%Type = Article
\bibitem[{de~Souza~Mendes et~al.(2007)de~Souza~Mendes, Naccache, Varges, and
  Marchesini}]{deSouzaMendes_07}
\bibinfo{author}{P.~R. de~Souza~Mendes}, \bibinfo{author}{M.~F. Naccache},
  \bibinfo{author}{P.~R. Varges}, \bibinfo{author}{F.~H. Marchesini},
\newblock \bibinfo{title}{Flow of viscoplastic liquids through axisymmetric
  expansions-contractions},
\newblock \bibinfo{journal}{Journal of Non-Newtonian Fluid Mechanics}
  \bibinfo{volume}{142} (\bibinfo{year}{2007}) \bibinfo{pages}{207 -- 217}.
%Type = Article
\bibitem[{Naccache and Barbosa(2007)}]{Naccache_07}
\bibinfo{author}{M.~F. Naccache}, \bibinfo{author}{R.~S. Barbosa},
\newblock \bibinfo{title}{Creeping flow of viscoplastic materials through a
  planar expansion followed by a contraction},
\newblock \bibinfo{journal}{Mechanics Research Communications}
  \bibinfo{volume}{34} (\bibinfo{year}{2007}) \bibinfo{pages}{423 -- 431}.
%Type = Article
\bibitem[{Botella and Peyret(1998)}]{Botella_98}
\bibinfo{author}{O.~Botella}, \bibinfo{author}{R.~Peyret},
\newblock \bibinfo{title}{Benchmark spectral results on the lid-driven cavity
  flow},
\newblock \bibinfo{journal}{Computers \& Fluids} \bibinfo{volume}{27}
  (\bibinfo{year}{1998}) \bibinfo{pages}{421 -- 433}.
%Type = Article
\bibitem[{Syrakos and Goulas(2006)}]{syrakos_06b}
\bibinfo{author}{A.~Syrakos}, \bibinfo{author}{A.~Goulas},
\newblock \bibinfo{title}{Finite volume adaptive solutions using {SIMPLE} as
  smoother},
\newblock \bibinfo{journal}{International Journal for Numerical Methods in
  Fluids} \bibinfo{volume}{52} (\bibinfo{year}{2006})
  \bibinfo{pages}{1215--1245}.
%Type = Article
\bibitem[{Bruneau and Saad(2006)}]{Bruneau_06}
\bibinfo{author}{C.-H. Bruneau}, \bibinfo{author}{M.~Saad},
\newblock \bibinfo{title}{The 2{D} lid-driven cavity problem revisited},
\newblock \bibinfo{journal}{Computers \& Fluids} \bibinfo{volume}{35}
  (\bibinfo{year}{2006}) \bibinfo{pages}{326 -- 348}.
%Type = Article
\bibitem[{Sanchez(1998)}]{Sanchez_98}
\bibinfo{author}{F.~Sanchez},
\newblock \bibinfo{title}{Application of a first-order operator splitting
  method to {B}ingham fluid flow simulation},
\newblock \bibinfo{journal}{Computers \& Mathematics with Applications}
  \bibinfo{volume}{36} (\bibinfo{year}{1998}) \bibinfo{pages}{71 -- 86}.
%Type = Article
\bibitem[{Mitsoulis and Zisis(2001)}]{mitsoulis_01}
\bibinfo{author}{E.~Mitsoulis}, \bibinfo{author}{T.~Zisis},
\newblock \bibinfo{title}{Flow of {B}ingham plastics in a lid-driven square
  cavity},
\newblock \bibinfo{journal}{Journal of Non-Newtonian Fluid Mechanics}
  \bibinfo{volume}{101} (\bibinfo{year}{2001}) \bibinfo{pages}{173--180}.
%Type = Article
\bibitem[{Vola et~al.(2003)Vola, Boscardin, and Latché}]{Vola_03}
\bibinfo{author}{D.~Vola}, \bibinfo{author}{L.~Boscardin},
  \bibinfo{author}{J.~Latché},
\newblock \bibinfo{title}{Laminar unsteady flows of {B}ingham fluids: a
  numerical strategy and some benchmark results},
\newblock \bibinfo{journal}{Journal of Computational Physics}
  \bibinfo{volume}{187} (\bibinfo{year}{2003}) \bibinfo{pages}{441 -- 456}.
%Type = Article
\bibitem[{Elias et~al.(2006)Elias, Martins, and Coutinho}]{Elias_06}
\bibinfo{author}{R.~Elias}, \bibinfo{author}{M.~Martins},
  \bibinfo{author}{A.~Coutinho},
\newblock \bibinfo{title}{Parallel edge-based solution of viscoplastic flows
  with the {SUPG/PSPG} formulation},
\newblock \bibinfo{journal}{Computational Mechanics} \bibinfo{volume}{38}
  (\bibinfo{year}{2006}) \bibinfo{pages}{365--381}.
%Type = Article
\bibitem[{Yu and Wachs(2007)}]{Yu_07}
\bibinfo{author}{Z.~Yu}, \bibinfo{author}{A.~Wachs},
\newblock \bibinfo{title}{A fictitious domain method for dynamic simulation of
  particle sedimentation in {B}ingham fluids},
\newblock \bibinfo{journal}{Journal of Non-Newtonian Fluid Mechanics}
  \bibinfo{volume}{145} (\bibinfo{year}{2007}) \bibinfo{pages}{78--91}.
%Type = Article
\bibitem[{Olshanskii(2009)}]{Olshanskii_09}
\bibinfo{author}{M.~A. Olshanskii},
\newblock \bibinfo{title}{Analysis of semi-staggered finite-difference method
  with application to {B}ingham flows},
\newblock \bibinfo{journal}{Computer Methods in Applied Mechanics and
  Engineering} \bibinfo{volume}{198} (\bibinfo{year}{2009}) \bibinfo{pages}{975
  -- 985}.
%Type = Article
\bibitem[{Zhang(2010)}]{Zhang_10}
\bibinfo{author}{J.~Zhang},
\newblock \bibinfo{title}{An augmented {L}agrangian approach to {B}ingham fluid
  flows in a lid-driven square cavity with piecewise linear equal-order finite
  elements},
\newblock \bibinfo{journal}{Computer Methods in Applied Mechanics and
  Engineering} \bibinfo{volume}{199} (\bibinfo{year}{2010})
  \bibinfo{pages}{3051 -- 3057}.
%Type = Article
\bibitem[{dos Santos et~al.(2011)dos Santos, Frey, Naccache, and
  de~Souza~Mendes}]{dosSantos_11}
\bibinfo{author}{D.~D. dos Santos}, \bibinfo{author}{S.~Frey},
  \bibinfo{author}{M.~F. Naccache}, \bibinfo{author}{P.~de~Souza~Mendes},
\newblock \bibinfo{title}{Numerical approximations for flow of viscoplastic
  fluids in a lid-driven cavity},
\newblock \bibinfo{journal}{Journal of Non-Newtonian Fluid Mechanics}
  \bibinfo{volume}{166} (\bibinfo{year}{2011}) \bibinfo{pages}{667 -- 679}.
%Type = Article
\bibitem[{Fedorenko(1962)}]{Fedorenko_62}
\bibinfo{author}{R.~P. Fedorenko},
\newblock \bibinfo{title}{A relaxation method for solving elliptic difference
  equations},
\newblock \bibinfo{journal}{USSR Computational Mathematics and Mathematical
  Physics} \bibinfo{volume}{1} (\bibinfo{year}{1962})
  \bibinfo{pages}{1092--1096}.
%Type = Article
\bibitem[{Brandt(1977)}]{Brandt_77}
\bibinfo{author}{A.~Brandt},
\newblock \bibinfo{title}{Multi-level adaptive solutions to boundary-value
  problems},
\newblock \bibinfo{journal}{Mathematics of Computation} \bibinfo{volume}{31}
  (\bibinfo{year}{1977}) \bibinfo{pages}{333--390}.
%Type = Article
\bibitem[{Sivaloganathan and Shaw(1988)}]{Sivaloganathan_88}
\bibinfo{author}{S.~Sivaloganathan}, \bibinfo{author}{G.~J. Shaw},
\newblock \bibinfo{title}{A multigrid method for recirculating flows},
\newblock \bibinfo{journal}{International Journal for Numerical Methods in
  Fluids} \bibinfo{volume}{8} (\bibinfo{year}{1988}) \bibinfo{pages}{417--440}.
%Type = Article
\bibitem[{Hortmann et~al.(1990)Hortmann, Peric, and G.}]{Hortmann_90}
\bibinfo{author}{M.~Hortmann}, \bibinfo{author}{M.~Peric},
  \bibinfo{author}{S.~G.},
\newblock \bibinfo{title}{Finite volume multigrid prediction of laminar natural
  convection: benchmark solutions},
\newblock \bibinfo{journal}{International Journal for Numerical Methods in
  Fluids} \bibinfo{volume}{11} (\bibinfo{year}{1990})
  \bibinfo{pages}{189--207}.
%Type = Article
\bibitem[{Tsamopoulos et~al.(1996)Tsamopoulos, Chen, and
  Borkar}]{tsamopoulos_96}
\bibinfo{author}{J.~A. Tsamopoulos}, \bibinfo{author}{M.~E. Chen},
  \bibinfo{author}{A.~V. Borkar},
\newblock \bibinfo{title}{On the spin coating of viscoplastic fluids},
\newblock \bibinfo{journal}{Rheologica Acta} \bibinfo{volume}{35}
  (\bibinfo{year}{1996}) \bibinfo{pages}{597--615}.
%Type = Article
\bibitem[{Burgos et~al.(1999)Burgos, Alexandrou, and Entov}]{burgos_99}
\bibinfo{author}{G.~R. Burgos}, \bibinfo{author}{A.~N. Alexandrou},
  \bibinfo{author}{V.~Entov},
\newblock \bibinfo{title}{On the determination of yield surfaces in
  {H}erschel-{B}ulkley fluids},
\newblock \bibinfo{journal}{Journal of Rheology} \bibinfo{volume}{43}
  (\bibinfo{year}{1999}) \bibinfo{pages}{463--483}.
%Type = Book
\bibitem[{Ferziger and Peric(2002)}]{Ferziger_02}
\bibinfo{author}{J.~H. Ferziger}, \bibinfo{author}{M.~Peric},
  \bibinfo{title}{Computational methods for fluid dynamics},
  \bibinfo{publisher}{Springer}, \bibinfo{edition}{3rd} edition,
  \bibinfo{year}{2002}.
%Type = Article
\bibitem[{Syrakos and Goulas(2006)}]{syrakos_06a}
\bibinfo{author}{A.~Syrakos}, \bibinfo{author}{A.~Goulas},
\newblock \bibinfo{title}{Estimate of the truncation error of finite volume
  discretization of the {N}avier-{S}tokes equations on colocated grids},
\newblock \bibinfo{journal}{International Journal for Numerical Methods in
  Fluids} \bibinfo{volume}{50} (\bibinfo{year}{2006})
  \bibinfo{pages}{103--130}.
%Type = Article
\bibitem[{Rhie and Chow(1983)}]{Rhie_Chow}
\bibinfo{author}{C.~M. Rhie}, \bibinfo{author}{W.~L. Chow},
\newblock \bibinfo{title}{Numerical study of the turbulent flow past an airfoil
  with trailing edge separation},
\newblock \bibinfo{journal}{AIAA Journal} \bibinfo{volume}{21}
  (\bibinfo{year}{1983}) \bibinfo{pages}{1525--1532}.
%Type = Book
\bibitem[{Patankar(1980)}]{Patankar_80}
\bibinfo{author}{S.~V. Patankar}, \bibinfo{title}{Numerical heat transfer and
  fluid flow}, \bibinfo{publisher}{Hemisphere}, \bibinfo{year}{1980}.
%Type = Book
\bibitem[{Saad(2003)}]{Saad}
\bibinfo{author}{Y.~Saad}, \bibinfo{title}{Iterative Methods for Sparse Linear
  Systems, 2nd edition}, \bibinfo{publisher}{SIAM}, \bibinfo{year}{2003}.
%Type = Article
\bibitem[{Shaw and Sivaloganathan(1988)}]{Shaw_88}
\bibinfo{author}{G.~J. Shaw}, \bibinfo{author}{S.~Sivaloganathan},
\newblock \bibinfo{title}{On the smoothing properties of the {SIMPLE}
  pressure-correction algorithm},
\newblock \bibinfo{journal}{International Journal for Numerical Methods in
  Fluids} \bibinfo{volume}{8} (\bibinfo{year}{1988}) \bibinfo{pages}{441--461}.
%Type = Article
\bibitem[{Burgos and Alexandrou(1999)}]{burgos_99b}
\bibinfo{author}{G.~R. Burgos}, \bibinfo{author}{A.~N. Alexandrou},
\newblock \bibinfo{title}{Flow development of {H}erschel-{B}ulkley fluids in a
  sudden three-dimensional square expansion},
\newblock \bibinfo{journal}{Journal of Rheology} \bibinfo{volume}{43}
  (\bibinfo{year}{1999}) \bibinfo{pages}{485--498}.

\end{thebibliography}






% \end{thebibliography}

\end{document}